\documentclass[twocolumn,showpacs,amsmath,amssymb,floats,floatfix,prb,aps,superscriptaddress,showkeys]{revtex4}

\usepackage{graphicx}
\usepackage{epsfig}
\usepackage{dcolumn}
\usepackage{bm}
\usepackage{times}

\begin{document}


\title{Kondo effect in single-molecule magnet transistors}

\author{Gabriel Gonz\'alez}

\affiliation{NanoScience Technology Center, University of Central
Florida, Orlando, FL 32826, USA,}

\affiliation{Department of Physics, University of Central
Florida, P.O. Box 162385, Orlando, FL 32816-2385, USA}

\author{Michael N. Leuenberger}\email{mleuenbe@mail.ucf.edu}

\affiliation{NanoScience Technology Center, University of Central
Florida, Orlando, FL 32826, USA,}

\affiliation{Department of Physics, University of Central
Florida, P.O. Box 162385, Orlando, FL 32816-2385, USA}

\author{Eduardo R. Mucciolo}\email{mucciolo@physics.ucf.edu}

\affiliation{Department of Physics, University of Central
Florida, P.O. Box 162385, Orlando, FL 32816-2385, USA}

\date{\today}

\begin{abstract}
We present a careful and thorough microscopic derivation of the Kondo
Hamiltonian for single-molecule magnets (SMMs) transistors. When the
molecule is strongly coupled to metallic leads, we show that by
applying a transverse magnetic field it is possible to topologically
induce or quench the Kondo effect in the conductance of a SMM with
either an integer or a half-integer spin $S>1/2$. This topological
Kondo effect is due to the Berry phase interference between multiple
quantum tunneling paths of the spin. We calculate the renormalized
Berry phase oscillations of the two Kondo peaks as a function of the
transverse magnetic field by means of the poor man's scaling. In
particular, we show that the Kondo exchange interaction between
itinerant electrons in the leads and the SMM pseudo spin 1/2 depends
crucially on the SMM spin selection rules for the addition and
subtraction of an electron and can range from antiferromagnetic to
ferromagnetic. We illustrate our findings with the SMM Ni$_4$, which
we propose as a possible candidate for the experimental observation of
the conductance oscillations.
\end{abstract}



\maketitle

\section{Introduction}
\label{sec:introduction}

Single-molecule magnets (SMMs), such as Mn$_{12}$ (see
Refs. \onlinecite{experiments,delBarco}) and Fe$_8$ (see
Refs. \onlinecite{Sangregorio,Wernsdorfer}), have become the focus of
intense research since experiments on bulk samples demonstrated the
quantum tunneling of a single magnetic moment on a macroscopic scale.
These molecules are characterized by a large total spin, a large
magnetic anisotropy barrier, and anisotropy terms which allow the spin
to tunnel through the barrier. Electronic transport through SMMs
offers several unique features with potentially large impact on
applications such as high-density magnetic storage as well as quantum
computing.\cite{app} Recent experiments have pointed out the
importance of the interference between spin tunneling paths in
molecules. For instance, measurements of the magnetization in bulk
Fe$_8$ have observed oscillations in the tunnel splitting
$\Delta_{m,-m}$ between states $S_z = m$ and $-m$ as a function of a
transverse magnetic field at temperatures between $0.05$ K and $0.7$ K
(see Ref. \onlinecite{Zener}). This effect can be explained by the
interference between Berry phases associated to spin tunneling paths
of opposite windings.\cite{LossDelftGarg,Leuenberger_Berry}
Theoretically, a coherent spin-state path integral formulation is used
to account for the coherence of the virtual states over which the spin
tunnels, although the initial and final spin states do not retain
their coherence.

A new approach to the study of SMMs opened up recently with the first
observation of quantized electronic transport through an {\it
isolated} Mn$_{12}$ molecule.\cite{heersche} One expects a rich
interplay between quantum tunneling, phase coherence, and electronic
correlations in the transport properties of SMMs. In fact, it has been
argued that the Kondo effect would only be observable for SMMs with
half-integer spin \cite{Romeike_Kondo,Wegewijs_Kondo} and therefore
absent for SMMs such as Mn$_{12}$, Fe$_8$, and Ni$_4$, where the spin
is integer. Later, two of us showed that this is not the
case:\cite{future} Remarkably, a transverse magnetic field $H_\bot$
can be tuned to topologically quench the two lowest levels of a
full-integer spin SMM, making them degenerate. In fact, the same
Berry-phase interference also affects transport for SMMs with
half-integer spin: In that case, sweeping $H_\bot$ will lead not to
one but {\it a series} of Kondo resonances. In the case of SMMs, as we
show below, the Berry phase oscillations of the tunnel splitting
$\Delta_{m_0,m'_0}$ of the uncharged single-molecule magnet ($q=0$)
for both full- and half-integer spins leads to oscillation of the
Kondo effect as a function of $H_\bot$. This means that the Kondo
effect is observable at zero bias for all values of the magnetic field
$H_{\bot,0}$ such that $\Delta_{m_0,m'_0}(H_{\bot,0})=0$.

It is interesting to note that at a finite bias the Kondo effect in a
quantum dot in the presence of a magnetic field can be restored by
tuning the bias to $eV = \pm g \mu_B H_\bot$ (see
Ref.~\onlinecite{Goldhaber}). For SMMs, however, the interference
between the Berry phases of the molecule total spin makes the distance
between the split Kondo peaks, which is equal to $eV = \pm
\Delta_{m_0,m'_0}$, oscillate as a function of $H_\bot$. A necessary
condition for observing these oscillations is a large enough tunnel
splitting.  Recently a new SMM based on tetranuclear nickel clusters
Ni$_4$ with a $S=4$ ground state has been synthesized. \cite{Sieber}
Our motivation for studying this particular nanomagnet stems partly
from its high symmetry $(S_{4})$ but, more importantly, from the large
tunnel splittings: $\Delta_{m_0,-m_0}\sim 0.01$ K or larger (depending
on the transverse field $H_\bot$) between the $|m\rangle=|4\rangle$
and $|m'\rangle=|-4\rangle$ ground states.

Recently, some authors have argued that the Kondo effect is absent at
the diabolic points of the Berry-phase
interference.\cite{Wegewijs_Kondo} This conclusion came from
considering an ad hoc Kondo Hamiltonian
\cite{Romeike_Kondo,Wegewijs_Kondo} which was not derived
microscopically. Moreover, in recent analysis of the sequential
tunneling regime, an ad hoc exchange Hamiltonian mixed with an
Anderson-type Hamiltonian was also used without a proper microscopic
derivation.\cite{Timm} In this paper, we provide a careful microscopic
derivation of the Kondo Hamiltonian suitable for full- and
half-integer spin single-molecule magnets by means of a
Schrieffer-Wolff transformation.\cite{SchriefferWolff} By using the
exact eigenstates of the positively and negatively charged
single-molecule magnet, it is sufficient to apply the Schrieffer-Wolff
transformation to second order in the tunneling matrix element. The
resulting Kondo exchange parameters exhibit the interference between
the second-order transition paths going over the two virtual charged
states. We show that this very same interference phenomenon is also
responsible for the Berry-phase blockade of the current through the
single-molecule magnet in the cotunneling regime, which extends our
previous results obtained for the Berry-phase blockade in the
sequential tunneling regime.\cite{gabriel}

Our derivation of the Kondo Hamiltonian reveals an important detail:
The Anderson-type Hamiltonian of the single-molecule magnet can be
mapped onto a spin-1/2 Kondo Hamiltonian in two different ways: Let
$S_{q=0}$ be the total spin of the uncharged single-molecule magnet.
\begin{enumerate}
\item If the total spins $S_{q=1}$ and $S_{q=-1}$ in the ground state
of the positively ($q=-1$) and negatively ($q=1$) charged
single-molecule magnet are equal to $S_{q=\pm 1}=S_{q=0}-1/2$, then
the Kondo Hamiltonian exhibits an {\it antiferromagnetic exchange
coupling}, which corresponds to the Kondo problem for spin 1/2
impurities or spin 1/2 quantum dots.
\item If $S_{q=\pm 1}=S_{q=0}+1/2$, then the Kondo Hamiltonian
exhibits a {\it ferromagnetic exchange coupling} that leads to a
vanishing renormalized transverse exchange coupling (Ising
interaction), in which case the Kondo effect is absent.\cite{hewson}
\end{enumerate}
This point is at the very heart of the Kondo problem. The Kondo effect
depends crucially on the selection rules for the addition and
subtraction of a spin on the molecule. This result is in contrast to
the Kondo effect seen in lateral quantum dots where the exchange
coupling is always antiferromagnetic due to fact that spin states are
degenerate in the absence of anisotropies.\cite{Glazman2005}

In the following, we provide a complete and detailed description of
the Kondo effect in SMMs. Starting from a microscopic model
(Sec. \ref{sec:model}), we derive the effective Kondo Hamiltonian for
a SMM attached to metallic leads through tunneling barriers
(Sec. \ref{sec:kondo}). In Sec. \ref{sec:results}, we derive
expressions for the conductance through a SMM for both zero and finite
bias as a function of a transverse magnetic field and use the new SMM
Ni$_4$ to comment the experimental significance of our theoretical
results. Our conclusions are summarized in Sec. \ref{sec:conclusions}.

\section{Microscopic Hamiltonian}
\label{sec:model}

\begin{figure}[ht]
\includegraphics[width=5cm]{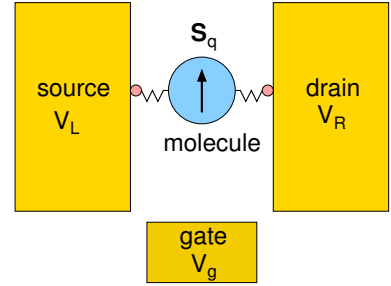}
\caption{(Color online) Schematic illustration of a single-molecule
field-effect transistor formed by a single-molecule magnet attached to
two metallic leads (source and drain) and controlled by a back gate
voltage.}
\label{fig:setup}
\end{figure}

The total Anderson-impurity-like Hamiltonian of a system formed by a
single-molecule magnet (SMM) attached to two metallic leads can be
separated into three terms (see Fig. \ref{fig:setup}),
\begin{equation}
\label{eq:Htotal}
{\cal H}_{\rm tot} = {\cal H}_{\rm SMM} + {\cal H}_{\rm lead} + {\cal
H}_{\rm SMM-lead}.
\end{equation}
The first term on the right-hand side of Eq. (\ref{eq:Htotal}) denotes
the SMM part, which can be broken into spin, orbital, charging, and
gate contributions,
\begin{equation}
\label{eq:HSMM}
{\cal H}_{\rm SMM} = {\cal H}_{\rm spin}^{(q)} + {\cal H}_{\rm
orbital} + \frac{q^2}{2}\, U - q\, e V_g,
\end{equation}
where $U$ denotes the charging energy, $q$ is the number of excess
electrons (the charge state of the molecule), and $V_g$ is the
electric potential due to an external gate voltage.\cite{obs1} In the
presence of an external magnetic field, the spin Hamiltonian of the
SMM reads
\begin{eqnarray}
{\cal H}_{\rm spin}^{(q)} & = & -A_qS_{q,z}^2+\frac{B_{2,q}}{2} \left(
S_{q,+}^2 + S_{q,-}^2 \right) \nonumber \\ & & + \frac{B_{4,q}}{3}
\left( S_{q,+}^4 + S_{q,-}^4 \right) \nonumber \\ & & + \frac{1}{2}
\left( h_\bot^\ast S_{q,+}+h_\bot S_{q,-} \right) + h_\parallel
S_{q,z},
\label{H_spin}
\end{eqnarray}
where the easy axis is taken along the $z$ direction and $S_{q,\pm} =
S_{q,x} \pm iS_{q,y}$. The magnetic field components were rescaled to
$h_\bot = g\mu_B (H_x + iH_y)$ and $h_\parallel = g\mu_B H_z$ for the
transversal and longitudinal parts, respectively, where $g$ denotes
the electron gyromagnetic factor. Note that the transverse magnetic
field lies in the $xy$ plane. In this Hamiltonian, the dominant
longitudinal anisotropy term creates a ladder structure in the
molecule spectrum where the $|\pm m_q\rangle$ eigenstates of $S_{q,z}$
are degenerate. The weak transverse anisotropy terms couple these
states. The total spin as well as the coupling parameters depend on
the charging state of the molecule. For example, it is known that
Mn$_{12}$ changes its easy-axis anisotropy constant (and its total
spin) from $A_0=56$ $\mu$eV ($S_{q=0}=10$) to $A_{-1}=43$ $\mu$eV
($S_{q=1}=19/2$) and $A_{-2}=32$ $\mu$eV ($S_{q=2}=10$) when singly and
doubly charged, respectively.\cite{dataexps}

The orbital contribution to the SMM energy is given by
\begin{equation}
{\cal H}_{\rm orbital} = \sum_{n,\sigma} \varepsilon_{n\sigma}
\psi^\dagger_{n\sigma} \psi_{n\sigma},
\end{equation}
where $\psi^\dagger_{n\sigma}$ ($\psi_{n\sigma}$) creates
(annihilates) electrons in the molecular orbital state $n$ with spin
orientation $\sigma$ and energy $\varepsilon_{n\sigma}$. Here we
neglect any diamagnetic response to external magnetic fields.

The second and third terms on the right-hand side of
Eq. (\ref{eq:Htotal}) read, respectively,
\begin{eqnarray}
\label{eq:Hlead}
{\cal H}_{\rm lead} & = & \sum_{a,k,\sigma} \left[ \xi_k^{(a)} +
\frac{1}{2} \left( h_\bot^\ast{\bf s}_{a,+} + h_\bot {\bf s}_{a,-}
\right) + h_\parallel {\bf s}_{a,z} \right] \nonumber \\ & & \times
\psi_{k\sigma,a}^\dagger \psi_{k\sigma,a}
\end{eqnarray}
and
\begin{equation}
\label{eq:SMM-lead}
{\cal H}_{\rm SMM-lead} = -\sum_{a,k,\sigma,n} \left[ t_{n,k}^{(a)}
\psi^\dagger_{k\sigma,a}\psi_{n\sigma} + {\rm H.c.} \right],
\end{equation}
where ${\bf s}_{a} = \sum_{k,k^\prime,\sigma,\sigma^\prime}
\sum_{a=R,L} \psi_{k\sigma,a}^\dagger ({\bf
\sigma}_{\sigma\sigma^\prime}/2) \psi_{k^\prime \sigma^\prime,a}$ and
$t_{n,k}^{(a)}$ is the lead-molecule tunneling amplitude. The operator
$\psi_{k\sigma,a}^\dagger$ ($\psi_{k\sigma,a}$) creates (annihilates)
electronic states in the $a$ lead ($a=R,L$) with linear momentum $k$,
spin orientation $\sigma$, and energy $\xi_{k}^{(a)}$.

\section{The Kondo Hamiltonian}
\label{sec:kondo}

The weak tunneling amplitudes between the leads and the molecule and
the large charging energy cause an effective exchange interaction
between electrons in the leads and the spin of the molecule. Although
this interaction at first glance seems to have the familiar Kondo
$s$-$d$ form, it is actually unusual because the transversal coupling
involves only a subspace of the spin of the molecule. Below, we use
perturbation theory to carefully derive an expression for the
effective Kondo Hamiltonian of a SMM.

We begin by considering only the SMM and SMM-lead terms in
Eq. (\ref{eq:Htotal}). We divide the Hilbert space into subspaces
corresponding to distinct charge sectors of the SMM. Using a block
matrix representation, we have
\begin{equation}
\label{eq:Hmatrix}
{\cal H}_{\rm SMM} + {\cal H}_{\rm SMM-lead} = \left(
\begin{array}{ccccc} \ddots & & & & \\ & {\cal H}_{q-1} & {\cal V} & 0 & \\ 
& {\cal V}^\dagger & {\cal H}_q & {\cal W} & \\ & 0 & {\cal W}^\dagger
& {\cal H}_{q+1} & \\ & & & & \ddots
\end{array} \right),
\end{equation}
where ${\cal H}_q$ is the SMM Hamiltonian for the charge sector $q$
[see Eq. (\ref{eq:HSMM})], while ${\cal V}$ and ${\cal W}$ represent
the lead-SMM tunneling Hamiltonian, Eq. (\ref{eq:SMM-lead}). For
certain values of the gate voltage, the charging energy is compensated
and Coulomb blockade is lifted. Away from these resonant points, there
is an energy gap of order $U$ between consecutive diagonal elements in
Eq. (\ref{eq:Hmatrix}). Since $U \gg |t_{n,k}|$, we can assume that
the off-diagonal elements ${\cal V}$ and ${\cal W}$ are small
perturbations and use a Schrieffer-Wolff transformation to decouple
distinct charge sectors up to terms of order
$O(|t_{n,k}|^2/U^2)$.\cite{SchriefferWolff}

For simplicity, let us consider the sectors $q=-1,0,+1$ only, where $q$ denotes the number of excess electrons, and
write\cite{obs2}
\begin{equation}
{\cal H} = {\cal A} + {\cal B},
\end{equation}
where
\begin{equation}
{\cal A} = \left( \begin{array}{ccc} {\cal H}_{-1} & 0 & 0 \\ 0 &
{\cal H}_{0} & 0 \\ 0 & 0 & {\cal H}_{+1} \end{array} \right), \ {\cal
B} = \left( \begin{array}{ccc} 0 & {\cal V} & 0 \\ {\cal V}^\dagger &
0 & {\cal W} \\ 0 & {\cal W}^\dagger & 0\end{array} \right).
\end{equation}
Using the similarity transformation $\tilde{\cal H} = e^{{\cal T}}
{\cal H}\, e^{-\cal T}$, where ${\cal T}$ is anti-Hermitian, we have
\begin{equation}
\tilde{\cal H} = {\cal H} + [{\cal T},{\cal H}] + \frac{1}{2} \left[
{\cal T}, [{\cal T},{\cal H}] \right] + \cdots.
\end{equation}
We want to determine ${\cal T}$ such that ${\cal B} + [{\cal T},{\cal
A}] = 0$. For that purpose, it is sufficient to assume that $T$ has the
form
\begin{equation}
{\cal T} = \left( \begin{array}{ccc} 0 & {\cal C} & 0 \\ -{\cal
C}^\dagger & 0 & {\cal D} \\ 0 & -{\cal D}^\dagger & 0 \end{array}
\right),
\end{equation}
with ${\cal C}$ and ${\cal D}$ satisfying
\begin{equation}
\label{eq:C}
{\cal H}_{-1} {\cal C} - {\cal C} {\cal H}_0 = {\cal V}
\end{equation}
and
\begin{equation}
\label{eq:D}
{\cal H}_0 {\cal D} - {\cal D} {\cal H}_{+1} = {\cal W},
\end{equation}
respectively. Thus,
\begin{equation}
\label{eq:Htilde}
\tilde{\cal H} = {\cal A} + \frac{1}{2} [{\cal T},{\cal B}] + O({\cal
B}^3),
\end{equation}
where
\begin{widetext}
\begin{equation}
\label{eq:TB}
[{\cal T},{\cal B}] = \left( \begin{array}{ccc} {\cal C} {\cal
V}^\dagger + {\cal V} {\cal C}^\dagger & 0 & {\cal C} {\cal W} - {\cal
V} {\cal D} \\ 0 & {\cal D} {\cal W}^\dagger + {\cal W} {\cal
D}^\dagger - {\cal C}^\dagger {\cal V} - {\cal V}^\dagger {\cal C} & 0
\\ {\cal W}^\dagger {\cal C}^\dagger - {\cal D}^\dagger {\cal
V}^\dagger & 0 & - {\cal D}^\dagger {\cal W} - {\cal W}^\dagger {\cal
D}.
\end{array} \right).
\end{equation}
\end{widetext}
Note that the neutral sector ($q=0$) has been decoupled from the
charged sectors at the expense of adding two contributions of order
$O({\cal V,W})^2$ to ${\cal H}_0$. To specify the form of these
contributions, we use the eigenbasis $\{|\alpha\rangle_q\}$ of ${\cal
H}_q$, namely,
\begin{equation}
{_q}\langle \alpha | {\cal H}_q | \beta \rangle_q = [{\cal
H}_q]_{\alpha\beta} = \delta_{\alpha,\beta}\, E^{(q)}_\alpha.
\end{equation}
Equations (\ref{eq:C}) and (\ref{eq:D}) can be solved in this
representation to yield
\begin{equation}
{_{-1}}\langle \alpha | {\cal C} | \beta \rangle_0 = [{\cal
C}]_{\alpha\beta} = \frac{[{\cal V}]_{\alpha\beta}} {E_\alpha^{(-1)} -
E_\beta^{(0)}}
\end{equation}
and
\begin{equation}
\label{eqD}
{_0}\langle \alpha | {\cal D} | \beta \rangle_{+1} = [{\cal
D}]_{\alpha\beta} = \frac{[{\cal W}]_{\alpha\beta}} {E_\alpha^{(0)} -
E_\beta^{(+1)}},
\end{equation}
respectively, where $[{\cal V}]_{\alpha\beta} = \left.{_{-1}}\langle
\alpha|{\cal V}|\beta \rangle_0\right.$ and $[{\cal W}]_{\alpha\beta}
= \left.{_0}\langle \alpha|{\cal W}|\beta \rangle_{-1}\right.$. This
allows us to write the following matrix elements for the neutral
sector:
\begin{equation}
\label{eq:DW}
{_0}\langle \alpha | {\cal D} {\cal W}^\dagger | \beta \rangle_0 =
\sum_{\gamma} \frac{[{\cal W}]_{\alpha\gamma} [{\cal
W}^\dagger]_{\gamma\beta}} {E_\alpha^{(0)} - E_{\gamma}^{(+1)}}
\end{equation}
and
\begin{equation}
\label{eq:VC}
{_0}\langle \alpha | {\cal V}^\dagger {\cal C} | \beta \rangle_0 =
\sum_{\gamma} \frac{[{\cal V}^\dagger]_{\alpha\gamma} [{\cal
V}]_{\gamma\beta}} {E_\gamma^{(-1)} - E_{\beta}^{(0)}}.
\end{equation}
%

\begin{figure}[t]
\includegraphics[width=8cm]{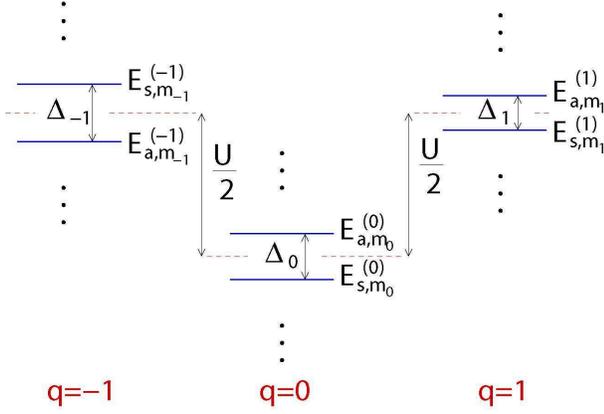}
\caption{(Color online) Energy spectra of consecutive charging sectors
of the SMM Hamiltonian ($q=-1,0,1$). $\Delta_q\equiv\Delta_{m_q,-m_q}$
is the tunnel splitting due to the in-plane anisotropy and
$E^{(q)}_{{\rm s/a},m_q}$ denote energy eigenstates corresponding to
symmetric and antisymmetric combinations of the ground eigenstates
$|\pm s_q\rangle$ of the longitudinal component of the SMM total spin
$S_{q,z}$.}
\label{fig:levels}
\end{figure}

In the Coulomb blockade valley, when $V_g = 0$ in Eq. (\ref{eq:HSMM}),
the eigenstates of the Hamiltonian ${\cal H}_q$ are expressed in terms
of symmetric and antisymmetric combinations of the eigenstates of the
$S_{q,z}$ operator, namely,
\begin{equation}
\label{eqS}
|{\rm s,a} \rangle_{m_q} = \frac{1}{\sqrt{2}} \left[ |m_q
 \rangle_0 \pm |-m_q\rangle_0 \right],
\end{equation}
with $m_q = 0,1,2, \ldots,S_q$ if $S_q$ is integer and $m_q =
\frac{1}{2},\frac{3}{2}, \ldots,S_q$ if $S_q$ is
half-integer.\cite{obs3} Note that since the matrices ${\cal W}$ and
${\cal V}$ represent the addition (subtraction) of an electron to
(from) the SMM, only $q=0$ states that differ in spin projection by
$1$ are coupled through Eqs. (\ref{eq:DW}) and
(\ref{eq:VC}). Conservation of angular momentum upon electron tunneling
requires that the intermediate states in the $q=1$ or $q=-1$ sectors
obey $|S_{\pm 1}-S_0|=1/2$. For instance, when the SMM total spin is
lowered by the addition or subtraction of an electron, longitudinal
spin components satisfy $m_{\pm 1}=S_{\pm 1}=S_{0}-1/2$ and $-m_{\pm
1}=-S_{\pm 1}=-S_{0}+1/2$. Therefore, we define the spin states of the
ground state to be $\left|\uparrow_q\right>=\left|m_q\right>$ and
$\left|\downarrow_q\right>=\left|-m_q\right>$.  The corresponding
eigenenergies are represented in Fig. \ref{fig:levels}.

Let us first consider intermediate states involving the $q=1$
sector. This situation corresponds to a small positive gate voltage
($V_g>0$). Equation (\ref{eq:DW}) can be used to write the matrix
elements of the reduced Hamiltonian of the $q=0$ sector in terms of a
product of energy denominators and matrix elements of ${\cal
W}$. Using the $\{|\uparrow\rangle_q,|\downarrow\rangle_q\}$ basis and
inserting complete eigenvector sets for each sector, i.e. ${\cal
I}^{(q)}=|s\rangle_{q} {_{q}}\langle s|+|a\rangle_{q} {_{q}}\langle
a|$, we find
\begin{widetext}
\begin{eqnarray}
\label{eq:uWWu}
{_0}\langle \uparrow| {\cal D} {\cal W}^\dagger + {\cal W} {\cal
  D}^\dagger | \uparrow \rangle_0 & = & {_0}\langle\uparrow|s\rangle_0
  {_0}\langle s| {\cal D}|s\rangle_1 {_1}\langle s| {\cal W}^\dagger
  |\uparrow\rangle_0 +\, {_0}\langle\uparrow|a\rangle_0 {_0}\langle a|
  {\cal D}|s\rangle_1 {_1}\langle s| {\cal W}^\dagger
  |\uparrow\rangle_0 \nonumber \\ & & \left. \ + \,
  {_0}\langle\uparrow|s\rangle_0 {_0}\langle s| {\cal D}|a\rangle_1
  {_1}\langle a| {\cal W}^\dagger |\uparrow\rangle_0 + \,
  {_0}\langle\uparrow|a\rangle_0 {_0}\langle a| {\cal D}|a\rangle_1
  {_1}\langle a| {\cal W}^\dagger |\uparrow\rangle_0 +\, \mbox{H.c.},
  \right.
\end{eqnarray}
\begin{eqnarray}
\label{eq:dWWd}
{_0}\langle \downarrow| {\cal D} {\cal W}^\dagger + {\cal W} {\cal
  D}^\dagger | \downarrow \rangle_0 & = &
  {_0}\langle\downarrow|s\rangle_0 {_0}\langle s| {\cal D}|s\rangle_1
  {_1}\langle s| {\cal W}^\dagger |\downarrow\rangle_0 +\,
  {_0}\langle\downarrow|a\rangle_0 {_0}\langle a| {\cal D}|s\rangle_1
  {_1}\langle s| {\cal W}^\dagger |\downarrow\rangle_0 \nonumber \\ &
  & \left. \ + \, {_0}\langle\downarrow|s\rangle_0 {_0}\langle s|
  {\cal D}|a\rangle_1 {_1}\langle a| {\cal W}^\dagger
  |\downarrow\rangle_0 + \, {_0}\langle\downarrow|a\rangle_0
  {_0}\langle a| {\cal D}|a\rangle_1 {_1}\langle a| {\cal W}^\dagger
  |\downarrow\rangle_0 +\, \mbox{H.c.}, \right.
\end{eqnarray}
and
\begin{eqnarray}
\label{eq:dWWu}
{_0}\langle \uparrow| {\cal D} {\cal W}^\dagger + {\cal W} {\cal
  D}^\dagger | \downarrow \rangle_0 & = & {_0}\langle \uparrow| {\cal
  D} {\cal W}^\dagger |\downarrow\rangle_0 + \left({_0}\langle
  \downarrow| {\cal D} {\cal W}^\dagger
  |\uparrow\rangle_0\right)^\dagger,
\end{eqnarray}
where
\begin{eqnarray}
\label{eq:dWWu1}
{_0}\langle \uparrow| {\cal D} {\cal W}^\dagger |\downarrow\rangle_0 &
= & {_0}\langle\uparrow|s\rangle_0 {_0}\langle s| {\cal D}|s\rangle_1
{_1}\langle s| {\cal W}^\dagger |\downarrow\rangle_0 +\,
{_0}\langle\uparrow|a\rangle_0 {_0}\langle a| {\cal D}|a\rangle_1
{_1}\langle a| {\cal W}^\dagger |\downarrow\rangle_0 \nonumber \\ & &
\left. \ + \, {_0}\langle\uparrow|s\rangle_0 {_0}\langle s| {\cal
D}|a\rangle_1 {_1}\langle a| {\cal W}^\dagger |\downarrow\rangle_0 +
\, {_0}\langle\uparrow|a\rangle_0 {_0}\langle a| {\cal D}|s\rangle_1
{_1}\langle s| {\cal W}^\dagger |\downarrow\rangle_0, \right.
\end{eqnarray}
and
\begin{eqnarray}
\label{eq:dWWu2}
{_0}\langle \downarrow| {\cal D} {\cal W}^\dagger |\uparrow\rangle_0 &
= & {_0}\langle\downarrow|s\rangle_0 {_0}\langle s| {\cal
D}|s\rangle_1 {_1}\langle s| {\cal W}^\dagger |\uparrow\rangle_0 +\,
{_0}\langle\downarrow|a\rangle_0 {_0}\langle a| {\cal D}|s\rangle_1
{_1}\langle s| {\cal W}^\dagger |\uparrow\rangle_0 \nonumber \\ & &
\left. \ + \, {_0}\langle\downarrow|s\rangle_0 {_0}\langle s| {\cal
D}|a\rangle_1 {_1}\langle a| {\cal W}^\dagger |\uparrow\rangle_0 + \,
{_0}\langle\downarrow|a\rangle_0 {_0}\langle a| {\cal D}|a\rangle_1
{_1}\langle a| {\cal W}^\dagger |\uparrow\rangle_0. \right.
\end{eqnarray}
Using Eq. (\ref{eqD}) and the fact that $E_{a}^{(+1)} =
E_{s}^{(+1)}+\Delta_1$, ${_0}\langle\uparrow|s\rangle_0 = \,
{_0}\langle\downarrow|s\rangle_0 = \, {_0}\langle\uparrow|a\rangle_0 =
1/\sqrt{2}$, and ${_0}\langle\downarrow|a\rangle_0 = -1/\sqrt{2}$, we
find, after some algebra,
\begin{eqnarray}
\label{eq1}
{_0}\langle \uparrow| {\cal D} {\cal W}^\dagger + {\cal W} {\cal
D}^\dagger | \uparrow \rangle_0 & = & \frac{1}{\sqrt{2}\Delta_s
\Delta_{sa}}\left[ {_0}\langle s|\left( \Delta_s {\cal W}{\cal
I}^{(1)}{\cal W}^\dagger - \Delta_1 {\cal W} |s\rangle_1 {_1}\langle
s|{\cal W}^\dagger \right)\ |\uparrow\rangle_0 \right] \nonumber \\ &
& \left. \ +\, \frac{1}{\sqrt{2}\Delta_a \Delta_{as}}\left[
{_0}\langle a|\left( \Delta_{as} {\cal W}{\cal I}^{(1)}{\cal
W}^\dagger - \Delta_1 {\cal W} |s\rangle_1 {_1}\langle s|{\cal
W}^\dagger \right) |\uparrow\rangle_0 \right] + \mbox{H.c.}, \right.
\end{eqnarray}
\begin{eqnarray}
\label{eq:dWWd2}
{_0}\langle \downarrow| {\cal D} {\cal W}^\dagger + {\cal W} {\cal
  D}^\dagger | \downarrow \rangle_0 & = & \frac{1}{\sqrt{2}\Delta_s
  \Delta_{sa}}\left[ _0\langle s|\left( \Delta_s {\cal W}{\cal
  I}^{(1)}{\cal W}^\dagger - \Delta_1 {\cal W} |s\rangle_1 {_1}\langle
  s|{\cal W}^\dagger \right)\ |\downarrow\rangle_0 \right] \nonumber
  \\ & & \left. \ -\, \frac{1}{\sqrt{2}\Delta_a \Delta_{as}}\left[
  {_0}\langle a|\left( \Delta_{as} {\cal W}{\cal I}^{(1)}{\cal
  W}^\dagger - \Delta_1 {\cal W} |s\rangle_1 {_1}\langle s|{\cal
  W}^\dagger \right) |\downarrow\rangle_0 \right] + \mbox{H.c.},
  \right.
\end{eqnarray}
\begin{eqnarray}
\label{eq:dWWu3}
{_0}\langle \uparrow| {\cal D} {\cal W}^\dagger |\downarrow\rangle_0 &
= & \frac{1}{\sqrt{2}\Delta_s \Delta_{sa}}\left[ {_0}\langle s|\left(
\Delta_s {\cal W}{\cal I}^{(1)}{\cal W}^\dagger - \Delta_1 {\cal W}
|s\rangle_1 {_1}\langle s|{\cal W}^\dagger \right)\
|\downarrow\rangle_0 \right] \nonumber \\ & & \left. \ +\,
\frac{1}{\sqrt{2}\Delta_a \Delta_{as}}\left[ {_0}\langle a|\left(
\Delta_{as} {\cal W}{\cal I}^{(1)}{\cal W}^\dagger - \Delta_1 {\cal W}
|s\rangle_1 {_1}\langle s|{\cal W}^\dagger \right)
|\downarrow\rangle_0 \right], \right.
\end{eqnarray}
and
\begin{eqnarray}
\label{eq:dWWu4}
{_0}\langle \downarrow| {\cal D} {\cal W}^\dagger |\uparrow\rangle_0 &
= & \frac{1}{\sqrt{2}\Delta_s \Delta_{sa}}\left[ {_0}\langle s|\left(
\Delta_s {\cal W}{\cal I}^{(1)}{\cal W}^\dagger - \Delta_1 {\cal W}
|s\rangle_1 {_1}\langle s|{\cal W}^\dagger \right)\ |\uparrow\rangle_0
\right] \nonumber \\ & & \left. \ -\, \frac{1}{\sqrt{2}\Delta_a
\Delta_{as}}\left[ {_0}\langle a|\left( \Delta_{as} {\cal W}{\cal
I}^{(1)}{\cal W}^\dagger - \Delta_1 {\cal W} |s\rangle_1 {_1}\langle
s|{\cal W}^\dagger \right) |\uparrow\rangle_0 \right], \right.
\end{eqnarray}
\end{widetext}
where $\Delta_s = E_s^{(0)}-E_s^{(+1)}$, $\Delta_{sa} =
E_s^{(0)}-E_a^{(+1)}$, $\Delta_a = E_a^{(0)}-E_a^{(+1)}$, and
$\Delta_{as} = E_a^{(0)}-E_s^{(+1)}$. To calculate the matrix elements
in Eqs. (\ref{eq1}), (\ref{eq:dWWd}), (\ref{eq:dWWu}), and
(\ref{eq:dWWu1}) we use the following definition for the operator
${\cal W}$ that originates from the lead-SMM Hamiltonian in
Eq. (\ref{eq:SMM-lead}):
\begin{equation}
{\cal W} = -\sum_{a,k,\sigma,n} t_{n,k}^{(a)}\,
\psi^\dagger_{k\sigma,a} \psi_{n\sigma},
\end{equation}
with $n$ being an occupied molecular orbital for a SMM in the charge
state $q=1$. Similarly, we define
\begin{equation}
{\cal W}^\dagger = \sum_{a,k,\sigma,n} t_{n,k}^{(a)}\,
\psi_{k\sigma,a} \psi_{n\sigma}^\dagger,
\end{equation}
which leads to
\begin{widetext}
\begin{equation}
{\cal W}{\cal I}^{(1)}{\cal W}^\dagger = -\sum_{ a,k,\sigma,n}
\sum_{a^\prime,k^\prime, \sigma^\prime, n^\prime} t_{n,k}^{(a)}
t_{n^\prime,k^\prime}^{(a^\prime)} \psi^\dagger_{k\sigma,a}
\psi_{k^\prime \sigma^\prime,a^\prime} \left( \psi_{n\sigma}{\cal
I}^{(1)} \psi_{n^\prime \sigma^\prime}^\dagger \right)
\label{eqwyw}
\end{equation}
and
\begin{equation}
{\cal W}|s\rangle_1 {_1}\langle s|{\cal W}^\dagger = -\sum_{
a,k,\sigma,n}\sum_{a^\prime,k^\prime,\sigma^\prime,n^\prime}
t_{n,k}^{(a)}t_{n^\prime,k^\prime}^{(a^\prime)}
\psi^\dagger_{k\sigma,a}\psi_{k^\prime \sigma^\prime,a^\prime}\left(
\psi_{n\sigma}|s\rangle_1 {_1}\langle s|\psi_{n^\prime
\sigma^\prime}^\dagger \right).
\label{eqwsw}
\end{equation}
\end{widetext}
%

\subsection{First set of spin selection rules}

We will now consider the case when adding or subtracting an electron
always {\it decreases} the SMM total spin, namely, $S_{q=\pm
1}=S_{q=0}-1/2$. This selection rule can be enforced through the
adoption of the following matrix elements:
\begin{equation}
{_0}\langle \uparrow | \psi_{n\sigma} | \uparrow \rangle_1 =
\delta_{m_0,m_1-\sigma}\, \delta_{\sigma,\downarrow},
\label{eqide1}
\end{equation}
\begin{equation}
{_0}\langle \downarrow | \psi_{n\sigma} | \downarrow \rangle_1  = 
\delta_{-m_0,-m_1-\sigma}\, \delta_{\sigma,\uparrow},
\label{eqide2}
\end{equation}
\begin{equation}
{_1}\langle \uparrow | \psi_{n\sigma}^\dagger | \uparrow \rangle_0 =
\delta_{m_0,m_1-\sigma}\, \delta_{\sigma,\downarrow},
\label{eqide3}
\end{equation}
and
\begin{equation}
{_1}\langle \downarrow | \psi_{n\sigma}^\dagger | \downarrow \rangle_0  = 
\delta_{-m_0,-m_1-\sigma}\, \delta_{\sigma,\uparrow}, 
\label{eqide4}
\end{equation}
(For the sake of simplicity, we will assume that $m_0,m_1 >
\frac{1}{2}$ and $\sigma=\pm \frac{1}{2}$ hereafter.) Then,
substituting Eq. (\ref{eqwyw}) and Eq. (\ref{eqwsw}) into
Eqs. (\ref{eq1}), (\ref{eq:dWWd}), (\ref{eq:dWWu}) and
(\ref{eq:dWWu1}), we arrive at
%

%
%
\begin{widetext}
\begin{equation}
\label{eq:uDW+WDu}
{_0}\langle\uparrow| {\cal D} {\cal W}^\dagger + {\cal W} {\cal
  D}^\dagger |\uparrow\rangle_0 = -\sum_{k,a,n}
  \sum_{k^\prime,a^\prime,n^\prime}
  J_{k,a;k^\prime,a^\prime}^{z\,(+1)} \psi^\dagger_{k\downarrow,a}
  \psi_{k^\prime \downarrow,a^\prime}\, - \sum_{k,a,n}
  \sum_{k^\prime,a^\prime,n^\prime} j_{k,a;k^\prime,a^\prime}^{\perp\,
  (+1)}\left( \psi^\dagger_{k\uparrow,a} \psi_{k^\prime
  \downarrow,a^\prime} + \psi^\dagger_{k\downarrow,a} \psi_{k^\prime
  \uparrow,a^\prime}\right),
\end{equation}
\begin{equation}
\label{eq:dDW+WDd}
{_0}\langle\downarrow| {\cal D} {\cal W}^\dagger + {\cal W} {\cal
  D}^\dagger |\downarrow\rangle_0 = -\sum_{k,a,n}
  \sum_{k^\prime,a^\prime,n^\prime}
  J_{k,a;k^\prime,a^\prime}^{z\,(+1)} \psi^\dagger_{k\uparrow,a}
  \psi_{k^\prime \uparrow,a^\prime}\, - \sum_{k,a,n}
  \sum_{k^\prime,a^\prime,n^\prime} j_{k,a;k^\prime,a^\prime}^{\perp\,
  (+1)}\left( \psi^\dagger_{k\uparrow,a} \psi_{k^\prime
  \downarrow,a^\prime} + \psi^\dagger_{k\downarrow,a} \psi_{k^\prime
  \uparrow,a^\prime}\right),
\end{equation}
\begin{eqnarray}
\label{eq:uDW+WDd}
{_0}\langle\uparrow| {\cal D} {\cal W}^\dagger + {\cal W} {\cal
 D}^\dagger |\downarrow\rangle_0 = -\sum_{k,a,n}
 \sum_{k^\prime,a^\prime,n^\prime} J_{k,a;k^\prime,a^\prime}^{\perp\,
 (+1)} \psi^\dagger_{k\downarrow,a} \psi_{k^\prime
 \uparrow,a^\prime}\, -\sum_{k,a,n} \sum_{k^\prime,a^\prime,n^\prime}
 j_{k,a;k^\prime,a^\prime}^{z\, (+1)}\left( \psi^\dagger_{k\uparrow,a}
 \psi_{k^\prime \uparrow,a^\prime} + \psi^\dagger_{k\downarrow,a}
 \psi_{k^\prime \downarrow,a^\prime}\right),
\end{eqnarray}
and
\begin{eqnarray}
\label{eq:dDW+WDu}
{_0}\langle\downarrow| {\cal D} {\cal W}^\dagger + {\cal W} {\cal
 D}^\dagger |\uparrow\rangle_0 = -\sum_{k,a,n}
 \sum_{k^\prime,a^\prime,n^\prime} J_{k,a;k^\prime,a^\prime}^{\perp\,
 (+1)} \psi^\dagger_{k\uparrow,a} \psi_{k^\prime
 \downarrow,a^\prime}\, -\sum_{k,a,n}
 \sum_{k^\prime,a^\prime,n^\prime} j_{k,a;k^\prime,a^\prime}^{z\,
 (+1)}\left( \psi^\dagger_{k\uparrow,a} \psi_{k^\prime
 \uparrow,a^\prime} + \psi^\dagger_{k\downarrow,a} \psi_{k^\prime
 \downarrow,a^\prime}\right).
\end{eqnarray}
Using a similar procedure, we also find 
\begin{equation}
\label{eqcv}
{_0}\langle\uparrow| {\cal C}^\dagger {\cal V} + {\cal V}^\dagger
  {\cal C} |\uparrow\rangle_0 = \sum_{k,a,n}
  \sum_{k^\prime,a^\prime,n^\prime}
  J_{k,a;k^\prime,a^\prime}^{z\,(-1)} \psi_{k^\prime
  \uparrow,a^\prime}\psi^\dagger_{k\uparrow,a}\, + \sum_{k,a,n}
  \sum_{k^\prime,a^\prime,n^\prime} j_{k,a;k^\prime,a^\prime}^{\perp\,
  (-1)}\left( \psi_{k^\prime
  \downarrow,a^\prime}\psi^\dagger_{k\uparrow,a} + \psi_{k^\prime
  \uparrow,a^\prime}\psi^\dagger_{k\downarrow,a}\right),
\end{equation}
\begin{equation}
\label{eqcv1}
{_0}\langle\downarrow| {\cal C}^\dagger {\cal V} + {\cal V}^\dagger
  {\cal C} |\downarrow\rangle_0 = \sum_{k,a,n}
  \sum_{k^\prime,a^\prime,n^\prime}
  J_{k,a;k^\prime,a^\prime}^{z\,(-1)} \psi_{k^\prime
  \downarrow,a^\prime}\psi^\dagger_{k\downarrow,a}\, + \sum_{k,a,n}
  \sum_{k^\prime,a^\prime,n^\prime} j_{k,a;k^\prime,a^\prime}^{\perp\,
  (-1)}\left( \psi_{k^\prime
  \downarrow,a^\prime}\psi^\dagger_{k\uparrow,a} + \psi_{k^\prime
  \uparrow,a^\prime}\psi^\dagger_{k\downarrow,a}\right),
\end{equation}
\begin{eqnarray}
\label{eqcv2}
{_0}\langle\uparrow| {\cal C}^\dagger {\cal V} + {\cal V}^\dagger
  {\cal C} |\downarrow\rangle_0 = \sum_{k,a,n}
  \sum_{k^\prime,a^\prime,n^\prime}
  J_{k,a;k^\prime,a^\prime}^{\perp\,(-1)} \psi_{k^\prime
  \uparrow,a^\prime}\psi^\dagger_{k\downarrow,a}\, + \sum_{k,a,n}
  \sum_{k^\prime,a^\prime,n^\prime} j_{k,a;k^\prime,a^\prime}^{z\,
  (-1)}\left( \psi_{k^\prime
  \uparrow,a^\prime}\psi^\dagger_{k\uparrow,a} + \psi_{k^\prime
  \downarrow,a^\prime}\psi^\dagger_{k\downarrow,a}\right),
\end{eqnarray}
and
\begin{eqnarray}
\label{eqcv3}
{_0}\langle\downarrow| {\cal C}^\dagger {\cal V} + {\cal V}^\dagger
  {\cal C} |\uparrow\rangle_0 = \sum_{k,a,n}
  \sum_{k^\prime,a^\prime,n^\prime} J_{k,a;k^\prime,a^\prime}^{\perp\,
  (-1)} \psi_{k^\prime
  \downarrow,a^\prime}\psi^\dagger_{k\uparrow,a}\, +\sum_{k,a,n}
  \sum_{k^\prime,a^\prime,n^\prime} j_{k,a;k^\prime,a^\prime}^{z\,
  (-1)}\left( \psi_{k^\prime
  \uparrow,a^\prime}\psi^\dagger_{k\uparrow,a} + \psi_{k^\prime
  \downarrow,a^\prime}\psi^\dagger_{k\downarrow,a}\right),
\end{eqnarray}
where
\begin{eqnarray}
J_{k,a;k^\prime,a^\prime}^{z\,(\pm 1)} & = & 2 t^{(a)}_{n,k}\,
t^{(a^\prime)}_{n^\prime,k^\prime}\,\left[
\frac{U+\Delta_0}{(U+\Delta_0)^2-\Delta_{\pm 1}^2}+ \frac{U -
\Delta_0}{(U-\Delta_0)^2-\Delta_{\pm 1}^2} \right]
\end{eqnarray}
\begin{eqnarray}
J_{k,a;k^\prime,a^\prime}^{\perp\,(\pm 1)} & = & \pm 2 t^{(a)}_{n,k}\,
t^{(a^\prime)}_{n^\prime,k^\prime}\,\left[ \frac{\Delta_{\pm 1}}{(U +
\Delta_0)^2 - \Delta_{\pm 1}^2} + \frac{\Delta_{\pm 1}}{(U -
\Delta_0)^2 - \Delta_{\pm 1}^2} \right],
\end{eqnarray}
\begin{eqnarray}
j_{k,a;k^\prime,a^\prime}^{z\,(\pm 1)} & = & t^{(a)}_{n,k}\,
t^{(a^\prime)}_{n^\prime,k^\prime}\,\left[
\frac{U+\Delta_0}{(U+\Delta_0)^2-\Delta_{\pm 1}^2} - \frac{U -
\Delta_0}{(U-\Delta_0)^2-\Delta_{\pm 1}^2} \right],
\end{eqnarray}
and
\begin{eqnarray}
j_{k,a;k^\prime,a^\prime}^{\perp\,(\pm 1)} & = & \pm t^{(a)}_{n,k}\,
t^{(a^\prime)}_{n^\prime,k^\prime}\,\left[ \frac{\Delta_{\pm 1}}{(U +
\Delta_0)^2 - \Delta_{\pm 1}^2} - \frac{\Delta_{\pm 1}}{(U -
\Delta_0)^2 - \Delta_{\pm 1}^2} \right].
\end{eqnarray}
\end{widetext}


In order to arrive at each matrix element expressed in
Eqs. (\ref{eq:uDW+WDu}) to (\ref{eq:dDW+WDu}) and Eqs. (\ref{eqcv}) to
(\ref{eqcv3}), we have only taken into account the leading terms in
the small ratio $\Delta_q/U$. For instance, in Eqs. (\ref{eq:uDW+WDu})
and (\ref{eq:dDW+WDd}) we have neglected spin-flipping terms which
carry an amplitude smaller than the direct, non spin-flipping terms by
$\Delta_0/U$. Note that the longitudinal exchange coupling is
positive, i.e. antiferromagnetic. Moreover, since $U \gg
\Delta_0,\Delta_{\pm 1}$, we find that the exchange couplings are
strongly anisotropic, with
\begin{equation}
|J^{\perp\,(\pm 1)}_{k,a;k^\prime,a^\prime}| \approx \frac{\Delta_{\pm
  1}}{U}\, J^{z\,(\pm 1)}_{k,a;k^\prime,a^\prime}.
\end{equation}
Finally, introducing pseudospin operators that act solely on the $q=0$
sector of the SMM, namely,
\begin{eqnarray}
\Sigma_0^z & = & |\uparrow\rangle_0\, {_0}\langle \uparrow|
- |\downarrow\rangle_0\, {_0}\langle \downarrow|, \\
\Sigma_0^+ & = & |\uparrow\rangle_0\, {_0}\langle
\downarrow|, \\
\Sigma_0^- & = & |\downarrow\rangle_0\, {_0}\langle \uparrow|,
\end{eqnarray}
and assuming, for brevity, that $\Delta_{1}=\Delta_{-1}$, we can write
the effective Hamiltonian of the $q=0$ charge sector as
\begin{widetext}
\begin{eqnarray}
\label{eq:KondoH}
\tilde{\cal H}_0 & = & {\cal H}_0 + 
\sum_{k^\prime,a^\prime}\sum_{k,a} \left[ J^z_{k,a;k^\prime,a^\prime}
\Sigma^z_{0} \left(  \psi^\dagger_{k\uparrow,a}
\psi_{k^\prime\uparrow,a^\prime} -\psi^\dagger_{k\downarrow,a}
\psi_{k^\prime\downarrow,a^\prime}\right) - 
J^\perp_{k,a;k^\prime,a^\prime} \left( \Sigma^+_{0}
\psi^\dagger_{k\downarrow,a} \psi_{k^\prime\uparrow,a^\prime} +
\Sigma^-_{0} \psi^\dagger_{k\uparrow,a}
\psi_{k^\prime\downarrow,a^\prime} \right)\right.  \nonumber \\ 
& & \left. \ -\,
j^\perp_{k,a;k^\prime,a^\prime}\left(
\psi^\dagger_{k\downarrow,a} \psi_{k^\prime\uparrow,a^\prime} +
\psi^\dagger_{k\uparrow,a} 
\psi_{k^\prime\downarrow,a^\prime} \right) \right],
\end{eqnarray}
where the last term in Eq. (\ref{eq:KondoH}) is a scattering term that
does not affect the dynamics of the SMM and can neglected. The
effective exchange coupling constants that appear in
Eq. (\ref{eq:KondoH}) are given by the following expressions:
\begin{eqnarray}
J^z_{k,a;k^\prime,a^\prime} &=& 2 t^{(a)}_{k}\,
t^{(a^\prime)}_{k^\prime}\,\left[
\frac{U+\Delta_0}{(U+\Delta_0)^2-\Delta_{1}^2}+ \frac{U -
\Delta_0}{(U-\Delta_0)^2-\Delta_{1}^2} \right], \label{eqjz+}\\
J^\perp_{k,a;k^\prime,a^\prime} &=& 4 t^{(a)}_{k}\,
t^{(a^\prime)}_{k^\prime}\,\left[ \frac{\Delta_{ 1}}{(U + \Delta_0)^2
- \Delta_{ 1}^2} + \frac{\Delta_{ 1}}{(U - \Delta_0)^2 - \Delta_{
1}^2} \right], \label{eqjperp} \\ j^\perp_{k,a;k^\prime,a^\prime} &=&
2 t^{(a)}_{k}\, t^{(a^\prime)}_{k^\prime}\,\left[ \frac{\Delta_{
1}}{(U + \Delta_0)^2 - \Delta_{ 1}^2} - \frac{\Delta_{1}}{(U -
\Delta_0)^2 - \Delta_{ 1}^2} \right].
\label{eqj}
\end{eqnarray}
\end{widetext}
The longitudinal exchange coupling is positive, which indicates an
antiferromagnetic Kondo exchange. We have neglected the dependence of
the hopping matrix elements on the SMM molecular orbital number
$n$. This is justified when the addition or subtraction of an electron
brings the molecule to the electronic ground state of the particular
charge sector. In this sense, only one orbital state can be filled
(emptied) when an electron is added (removed).

\begin{figure}[ht]
\includegraphics[width=7.6cm]{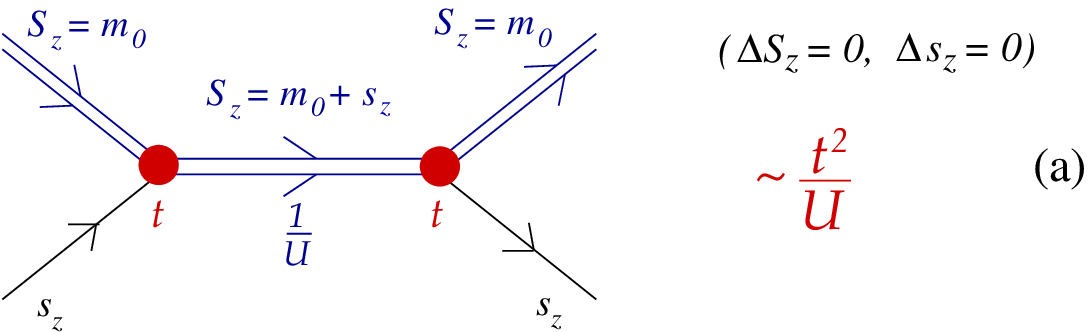}
\vspace{1cm}
\includegraphics[width=8cm]{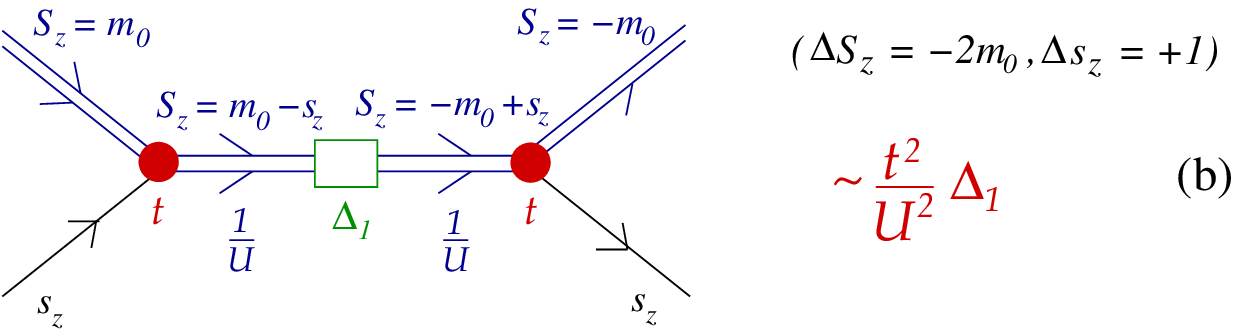}
\caption{(Color online) Diagrams representing the (a) longitudinal and
(b) transversal exchange interactions between the SMM magnetization
and the electrons in the leads.}
\label{fig:diagrams}
\end{figure}

A diagrammatic representation of the longitudinal and transverse
exchanging interactions is shown in Fig. \ref{fig:diagrams}. These
diagrams differ from the usual Kondo effect in the sense that the
transverse, spin-flipping interaction in a SMM requires a quantum
tunneling of the {\it total} magnetization during its virtual state.

\subsection{Second set of spin selection rules}

If we consider the case where adding or subtracting an electron always
{\it increases} the total spin in the SMM, namely, $S_{q=\pm
1}=S_{q=0}+1/2$, we have to modify Eqs. (\ref{eqide1}),
(\ref{eqide2}), (\ref{eqide3}), and (\ref{eqide1}) by adopting instead
the following matrix elements:
\begin{equation}
{_0}\langle \uparrow | \psi_{n\sigma} | \uparrow \rangle_1 =
\delta_{m_0,m_1+\sigma}\delta_{\sigma \uparrow},
\label{eqide1f}
\end{equation}
\begin{equation}
{_0}\langle \downarrow | \psi_{n\sigma} | \downarrow \rangle_1  = 
\delta_{-m_0,-m_1+\sigma}\delta_{\sigma \downarrow},
\label{eqide2f}
\end{equation}
\begin{equation}
{_1}\langle \uparrow | \psi_{n\sigma}^\dagger | \uparrow \rangle_0 =
\delta_{m_0,m_1+\sigma}\delta_{\sigma \uparrow},
\label{eqide3f}
\end{equation}
and
\begin{equation}
{_1}\langle \downarrow | \psi_{n\sigma}^\dagger | \downarrow \rangle_0  = 
\delta_{-m_0,-m_1+\sigma}\delta_{\sigma \downarrow}. 
\label{eqide4f}
\end{equation}
Using these selection rules, we arrive at a Kondo Hamiltonian with
exactly the same form as that in Eq. (\ref{eq:KondoH}). The
expressions for the exchange coupling constants are the same as those
given in Eqs. (\ref{eqjz+}), (\ref{eqjperp}), and (\ref{eqj}), except
for the longitudinal coupling $J^z_{k,a;k^\prime,a^\prime}$, which
{\it changes its overall sign and becomes negative}, signaling a
ferromagnetic Kondo exchange interaction. However, the strong
anisotropy remains, with the bare longitudinal coupling dominant over
the transversal one.

We can summarize our results so far by stating that
\begin{enumerate} 
\item if $S_{q=\pm 1}=S_{q=0}-1/2$ then $J^z_{k,a;k^\prime,a^\prime}
\gg J^\perp_{k,a;k^\prime,a^\prime} > 0$ (antiferromagnetic exchange
coupling);
\item if $S_{q=\pm 1}=S_{q=0}+1/2$ then $J^z_{k,a;k^\prime,a^\prime} \ll
  -J^\perp_{k,a;k^\prime,a^\prime} < 0$ (ferromagnetic exchange
  coupling).
\end{enumerate}

\section{Conductance and the Kondo effect in a SMM}
\label{sec:results}

In order to evaluate the conductance of the SMM subjected to the
Hamiltonian of Eq. (\ref{eq:KondoH}), we make use of the standard poor
man's scaling to renormalize the effective exchange coupling constants
$J^z$ and $J^\perp$ and the $g$-factor. We start by calculating the
renormalization flow at the points where the Kondo effect is
observable at zero bias, namely, where the tunnel splitting vanishes:
$\Delta_{0} \left( h_\bot^{(l)} \right) = 0$. For half-integer spins,
it is reasonable to assume that $h_\bot^{(l)} \gg T_K$, except for the
first zero ($l=0$). The total Hamiltonian reads
\begin{eqnarray}
\label{eq:Htot}
{\cal H}_{\rm tot} & = & \sum_m \left[ \epsilon_{m} + \frac{1}{2}\eta
(h_\bot^\ast \Sigma_+ + h_\bot \Sigma_-)\right] \nonumber \\ & & +
\sum_{k,s} \xi_k \psi_{ks} \psi_{ks}+\tilde{\cal H}_0 ,
\end{eqnarray}
where $\epsilon_{m}$ is the eigenvalue of $\left|m\right>$ for
$h_\bot=0$ and, due the Knight shift, $\eta = 1 - \rho_0 J^\perp/2$,
with $\rho_0$ denoting the density of states of the itinerant
electrons at the Fermi energy.\cite{Glazman2005} We do not include the
Zeeman term for the itinerant electrons in Eq. (\ref{eq:Htot}) because
at finite values of $h_\bot^{(l)}$ one has to cut the edges of the
spin-up and spin-down bands in the leads to make them symmetric with
respect to the Fermi energy.\cite{Glazman2005} We call $D$ the
resulting band width.

The Hamiltonian of Eq. (\ref{eq:Htot}) remains invariant under
renormalization group transformations (see the appendix for
details). Using Eq. (\ref{eqa11}), we obtain the flow equations
\begin{equation}
\label{jt}
\frac{dJ^\perp}{d\zeta} = 2\rho_0\, J^\perp J^z,
\end{equation}
\begin{equation}
\frac{dJ^z}{d\zeta} = 2\rho_0\, \left(J^\perp\right)^2,
\label{jz}
\end{equation}
and
\begin{equation}
\frac{d\eta}{d\zeta} =  - \rho_0^2\,J^\perp J^z,
\end{equation}
where $\zeta = \ln(\tilde{D}/D)$ and $\tilde{D}$ is the rescaled band
width. Dividing Eq. (\ref{jt}) by (\ref{jz}) and integrating by parts
gives $\left( J^z \right)^2 - \left( J^\perp \right)^2 = C^2$, where
$C$ is a positive constant.\cite{Anderson}

\begin{figure}[htb]
\includegraphics[width=8cm]{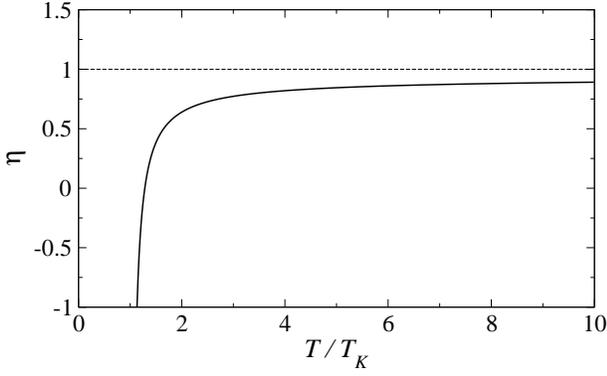}
\caption[]{The renormalization of the $g$-factor due to the Knight
shift. As an estimate, we use $\rho_0 J_\pm=\rho_0J_z=0.15$ where $\rho_0=9.45 \times 10^{20}$ J$^{-1}$ (see Ref. \onlinecite{Park}).}
\label{Eta}
\end{figure}

We have to distinguish between two cases: $J^z$ is either positive or
negative. If $J^z$ is positive, the exchange coupling constants remain
antiferromagnetic during the flow but the exchange interaction become
increasingly isotropic. Solving Eqs. (\ref{jt}) and (\ref{jz}) yields
\begin{equation}
\label{eq:flowsol}
\frac{1}{2\rho_0 C}\, {\rm arctanh} \left( \frac{C} {J^z} \right) =
\ln \left( \frac{\tilde{D}}{T_K} \right).
\end{equation}
The solution for $J^\perp$ is determined by $J^z = \sqrt{
\left(J^\perp\right)^2 + C^2}$. The flow of $\eta$ is shown in
Fig. \ref{Eta}. The flow stops at $\tilde{D}\approx \omega \sim T >
T_K$. In the antiferromagnetic case the Berry-phase oscillations get
strongly renormalized by the scaling of the Knight shift.

Since $\left|J^z\right| \gg J^\perp$, when $J^z$ is negative, the
transverse exchange coupling $J^\perp$ renormalizes to zero.
Therefore, in this case the Kondo resonance cannot form and the
interaction becomes Ising-like. The interesting feature of $J^\perp=0$
is that the Knight shift vanishes.

\subsection{Linear conductance}
\label{sec:linearG}

In order to calculate the linear conductance through the SMM we use
the following well-known expression for the weak coupling regime ($T_K
\ll T$):\cite{Glazman2005}
\begin{equation} 
G(T) = G_0 \int_{-\infty}^{\infty} d\omega\, \left(
-\frac{df}{d\omega} \right) \frac{\pi^2 \rho_0^2}{16}
\left|A(\omega)\right|^2,
\label{conductance}
\end{equation}
where $G_0$ is the classical (incoherent) conductance of the molecule,
$df/d\omega$ is the derivative of the Fermi function, and $A(\omega)$
is the transition amplitude. At the end of the flow, the transition
amplitude can be calculated in first-order perturbation theory as
\begin{equation} 
A_{\tilde{D}\approx\omega} = J^\perp_{\omega} = C \left[
\frac{(\omega/T_K)^{2\rho_0 C}} {(\omega/T_K)^{4\rho_0 C}-1} \right],
\label{amplitude} 
\end{equation}
The Knight shift is related to the transition amplitude by
\begin{equation}
\eta_{\tilde{D}\approx\omega} = 1 - \frac{\rho_0\, A_{\tilde{D}
\approx \omega}}{2}.
\label{eta}
\end{equation}
By making the substitution $\omega \rightarrow T$ into
Eq. (\ref{amplitude}), one finds that the linear conductance diverges
when $T \rightarrow T_K$, signaling the onset of the Kondo
effect. Since $C>0$, the singularity in Eq. (\ref{amplitude}) differs
from the usual logarithmic behavior found for isotropic exchange
interactions. We note, however, that in reality, the conductance does
not diverge but is rather strongly enhanced near the Kondo
temperature. The poor man's scaling breaks down near the Kondo
temperature and more accurate nonperturbative methods, such as the
density matrix renormalization group,\cite{DMRG} have to be employed
for obtaining a quantitative description of the conductance dependence
on temperature.

Using Eq. (\ref{conductance}), we get for the linear conductance
\begin{equation} 
\frac{G(T)}{G_0} = \frac{\pi^2\rho_0^2}{16} \left(
J^\perp_{\tilde{D}\approx T}\right)^2,
\end{equation}
which has the same functional form as the result of
Ref. \onlinecite{appelbaum} for the resistivity of bulk metals in the
presence of Kondo impurities.

\begin{figure}[htb]
  \begin{center}
    \begin{tabular}{cc}
      \resizebox{40mm}{!}{\includegraphics{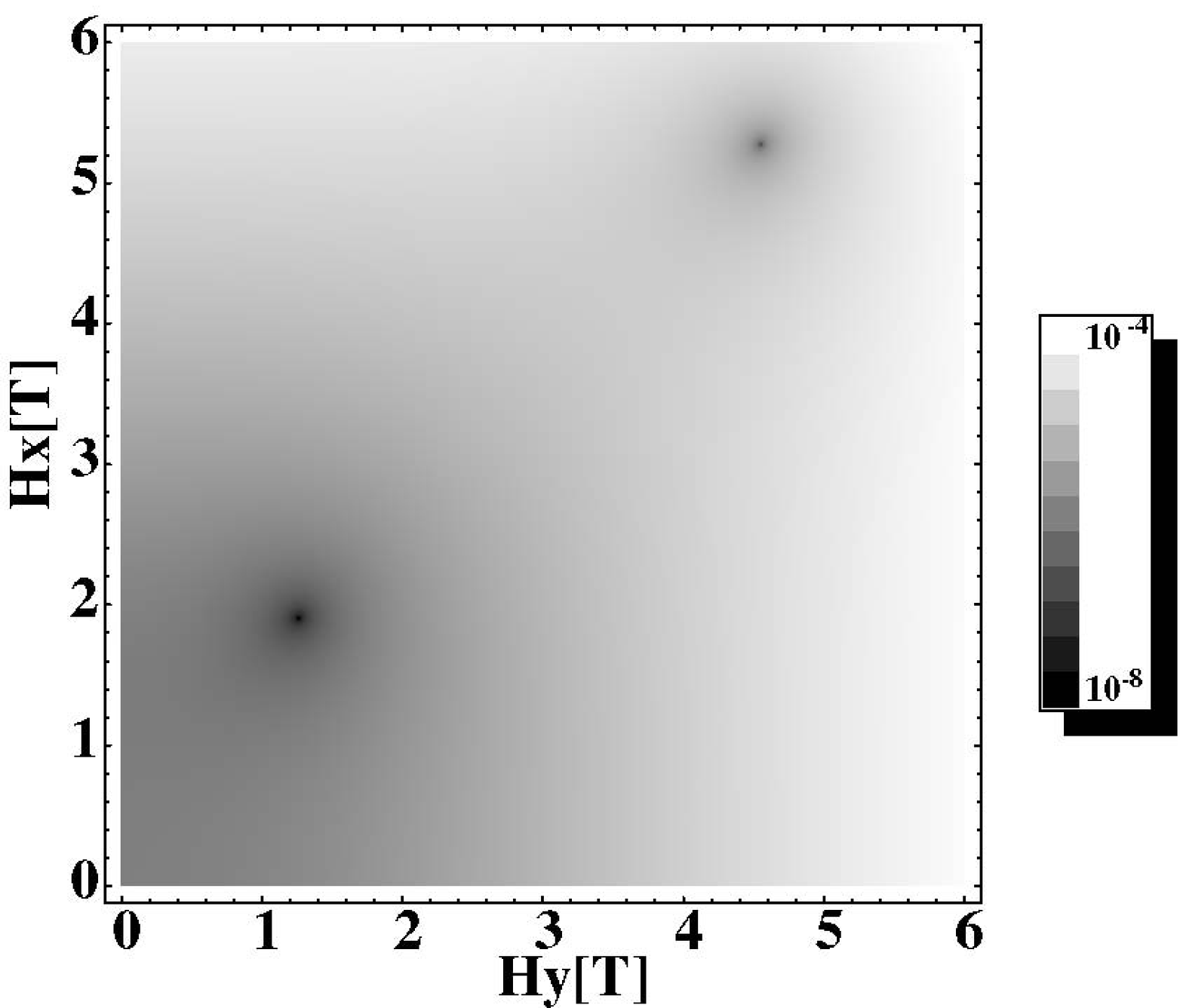}} &
      \resizebox{40mm}{!}{\includegraphics{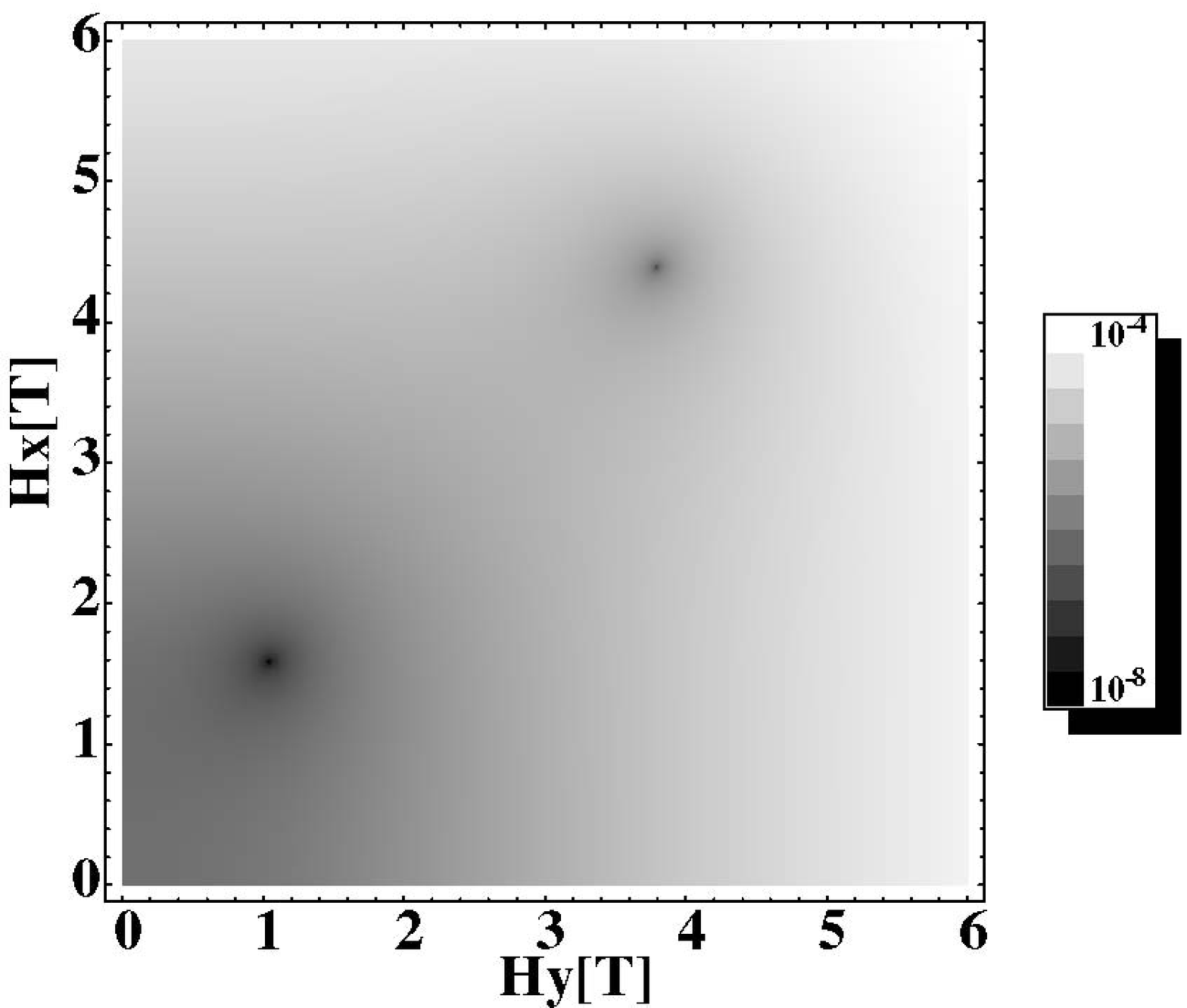}} \\ 
    \multicolumn{1}{c}{\mbox{\bf (a)}} &
		\multicolumn{1}{c}{\mbox{\bf (b)}} \\ 
      \resizebox{40mm}{!}{\includegraphics{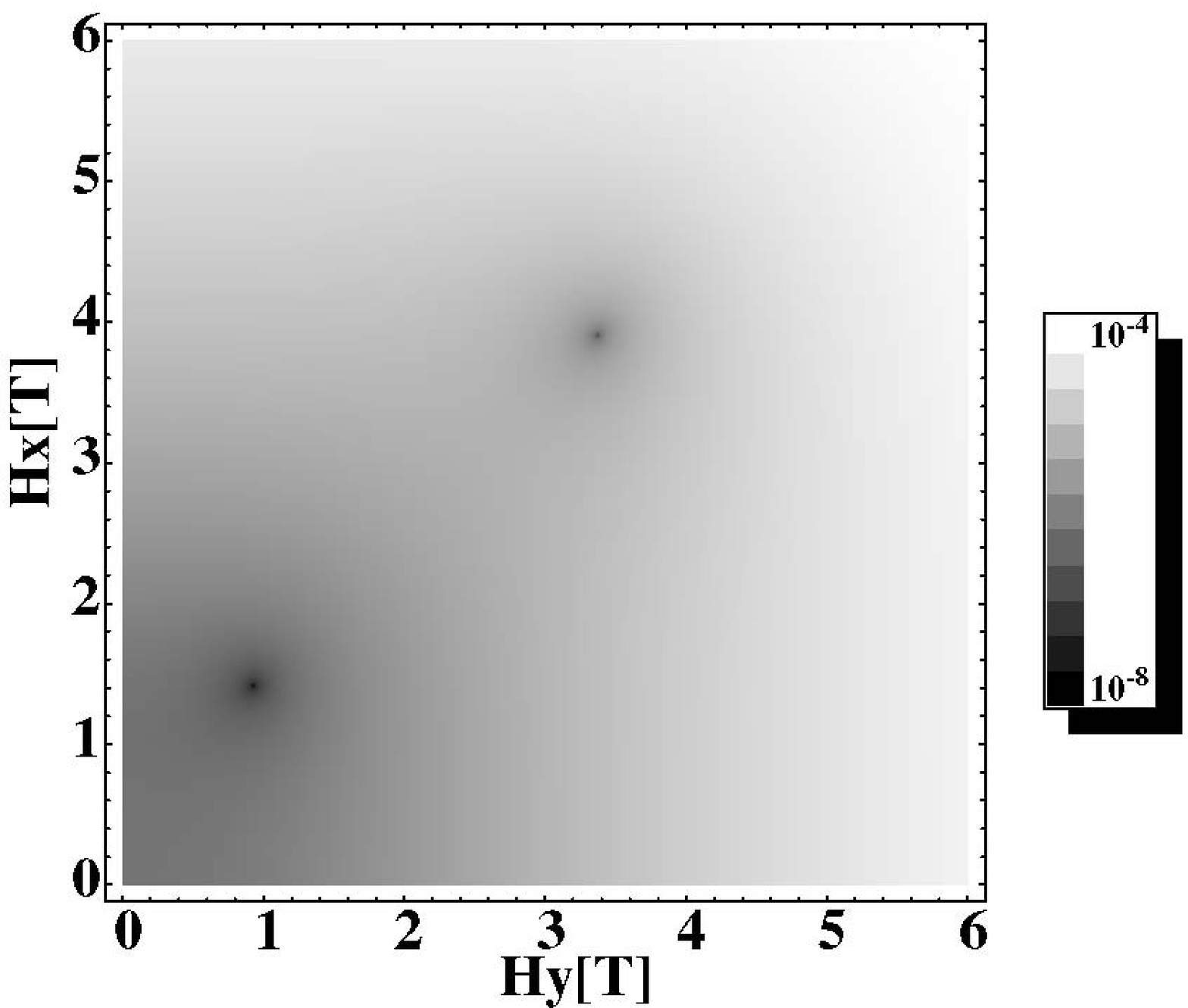}} &
      \resizebox{40mm}{!}{\includegraphics{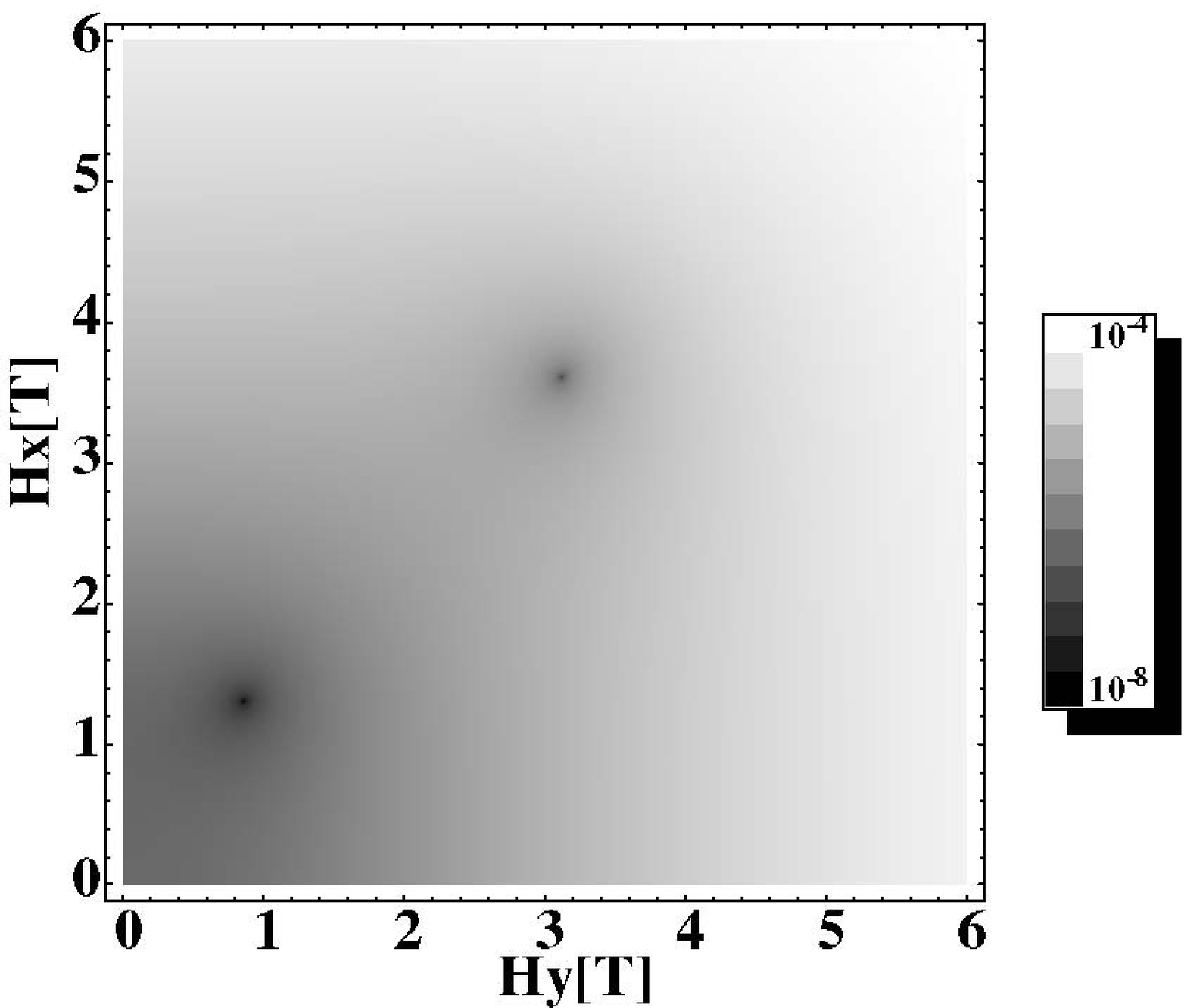}} \\
    \multicolumn{1}{c}{\mbox{\bf (c)}} &
		\multicolumn{1}{c}{\mbox{\bf (d)}} \\
      \resizebox{40mm}{!}{\includegraphics{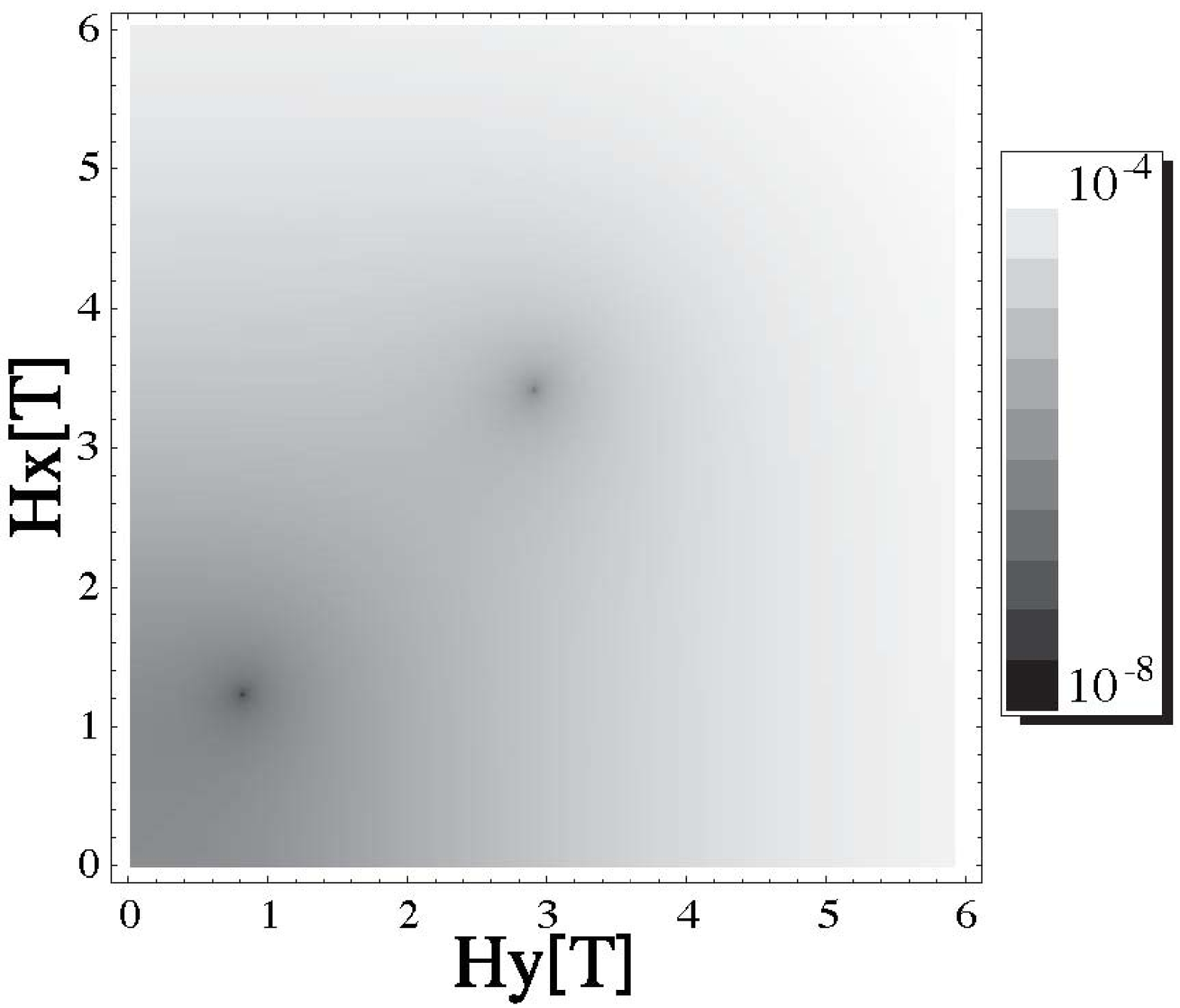}} &
      \resizebox{40mm}{!}{\includegraphics{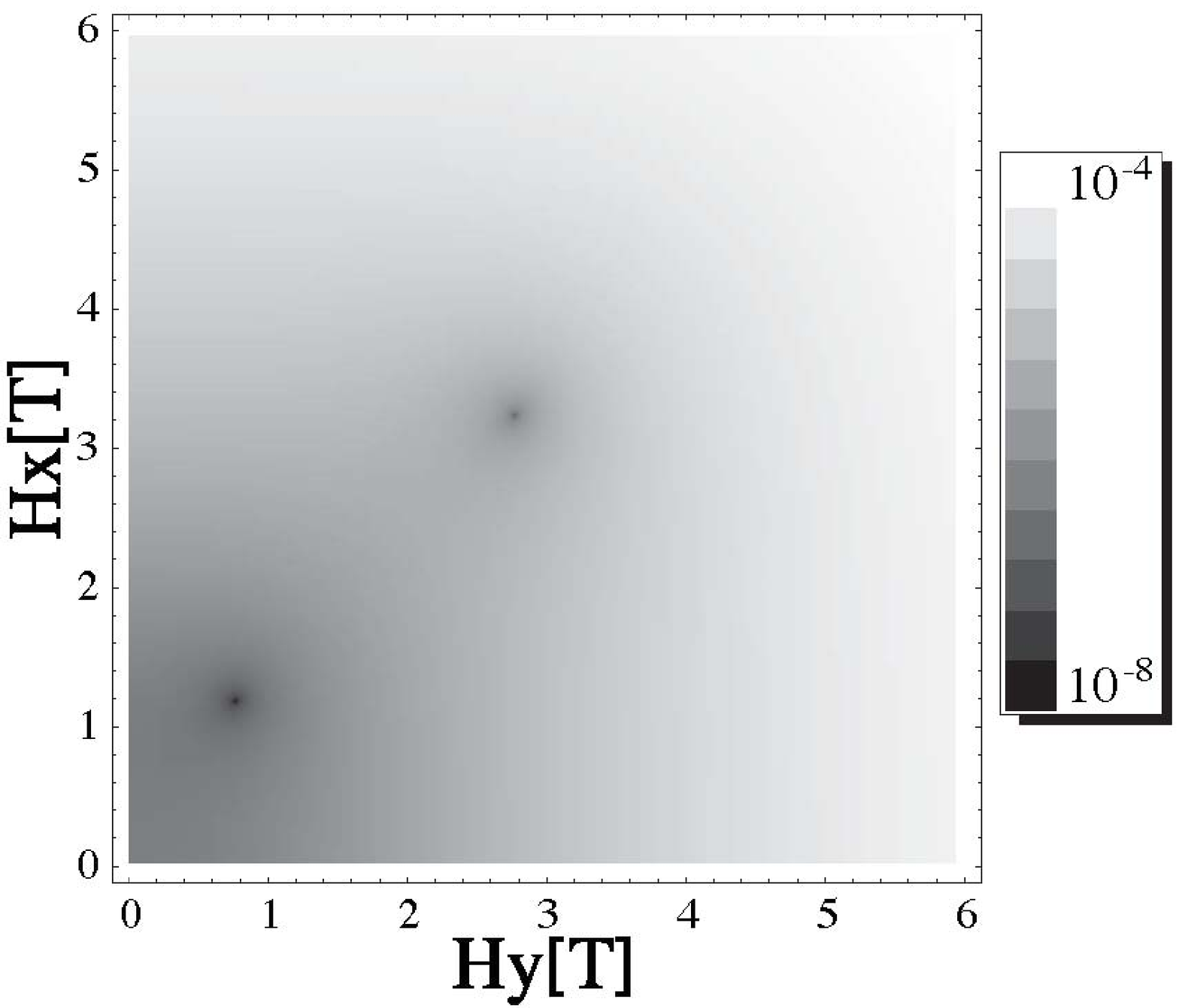}} \\
    \multicolumn{1}{c}{\mbox{\bf (e)}} &
		\multicolumn{1}{c}{\mbox{\bf (f)}} \\ 
    \end{tabular}
    \caption{The graph shows the temperature dependence of the zeros
    for the Berry phase oscillation as a function of the transverse
    magnetic field for the tunnel splittings between $|4\rangle$ and
    $|-4\rangle$ states ($S_0=4$) and the following temperature
    values: (a) $T/T_K=1.5$, (b) $T/T_K=1.6$, (c) $T/T_K=1.7$, (d)
    $T/T_K=1.8$, (e) $T/T_K=1.9$, and (f) $T/T_K=2.0$.}
	\label{BPBT}
  \end{center}
\end{figure}

All the zero points of the Berry phase oscillation are rescaled by the
$g$-factor renormalization: $b_\bot^{(l)} =
h_\bot^{(l)}/\eta_{\tilde{D} \approx T}$. Thus, the zero points become
dependent on the contributing states $|m\rangle$ and
$|-m\rangle$. This result indicates that the period of the Berry phase
oscillations becomes temperature dependent at $T>T_K$ (see
Fig. \ref{BPBT}). This fact allows us to conclude that the scaling
equations can be checked experimentally by measuring the renormalized
zero points of the Berry phase. Furthermore, due to the scale
invariance of the Kondo effect, the distance between the zeros should
follow a universal function of $T/T_K$ (see Fig. \ref{BPBT}).

\subsection{Nonlinear conductance}
\label{sec:nonlinearG}

Let us now study the conductance for nonzero bias, $V\neq 0$. For this
purpose, we first express the current flowing through the SMM in terms
of tunneling rates and lead occupation factors, namely,
\begin{equation}
I = \frac{e}{h}\sum_{\sigma}\int^{\infty}_0 dE\,
\frac{\Gamma_{L\sigma} \Gamma_{R\sigma}} {\Gamma_{L\sigma} +
\Gamma_{R\sigma}}\, \rho(E)\, \left[ f_{L}(E)-f_{R}(E) \right],
\label{current}
\end{equation}
where $\rho(E)$ is the energy dependent density of states,
$\Gamma_{L\sigma}$ ($\Gamma_{R\sigma}$) is the escape rate for the
left (right) lead, and $f_{L}$ ($f_{R}$) is the Fermi function for the
left (right) lead. For the sake of simplicity, we assume
$\Gamma_{L\sigma} = \Gamma_{R\sigma} = \Gamma$. Since, at low
temperatures,
\begin{equation}
f_{L}(E) - f_{R}(E) \approx\left\{ \begin{array}{ll} 1 & \mbox{if}\
E_F -eV/2 \leq E \leq E_F + eV/2, \\ 0 & \mbox {otherwise},
\end{array} \right.
\label{fermi}
\end{equation}
where $E_F$ is the Fermi energy of the leads, we get the following
expression for the differential conductance:
\begin{eqnarray}
G = \frac{dI}{dV} & = & \frac{\pi e^2}{2h} \int^{\infty}_0\, dE\,
|A(E)|^2\left[ \delta (E-E_F-eV/2) \right. \nonumber \\ & & \left. -\
  \delta (E-E_F+eV/2) \right].
\label{conductance1}
\end{eqnarray}
%

\begin{figure}[ht]
  \begin{center}
    \begin{tabular}{cc}
      \resizebox{40mm}{!}{\includegraphics{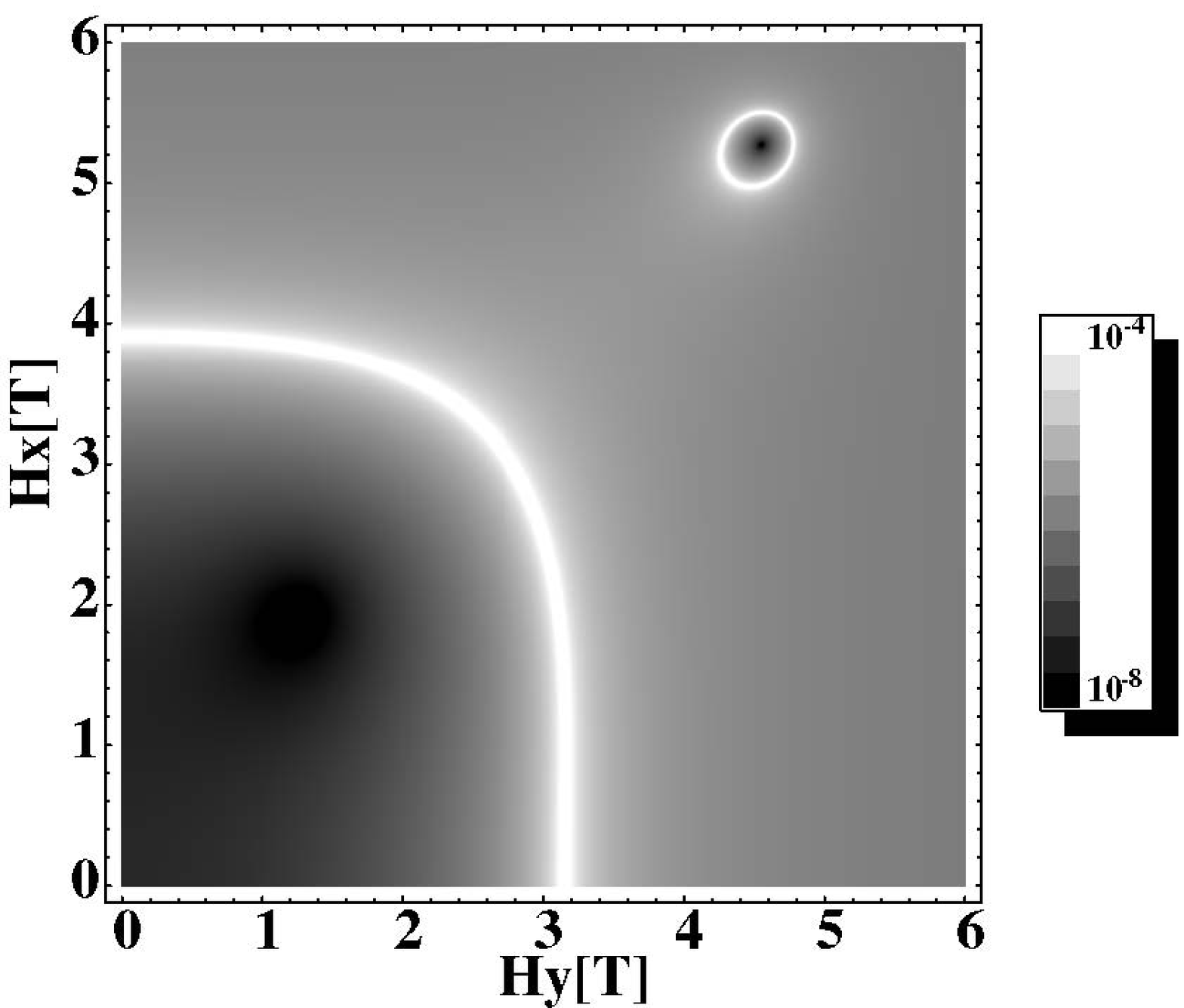}} &
      \resizebox{40mm}{!}{\includegraphics{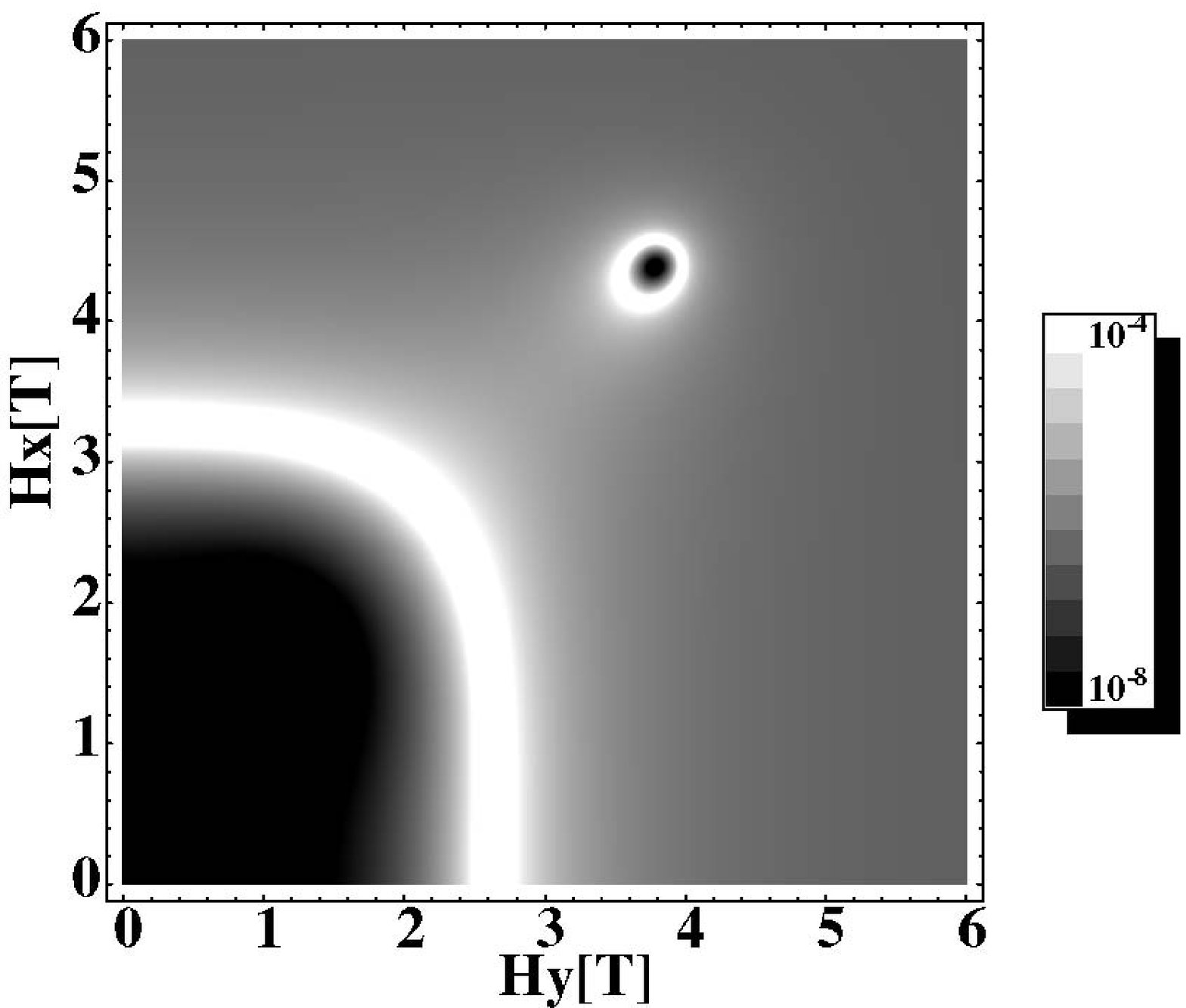}} \\ 
    \multicolumn{1}{c}{\mbox{\bf (a)}} &
		\multicolumn{1}{c}{\mbox{\bf (b)}} \\ 
      \resizebox{40mm}{!}{\includegraphics{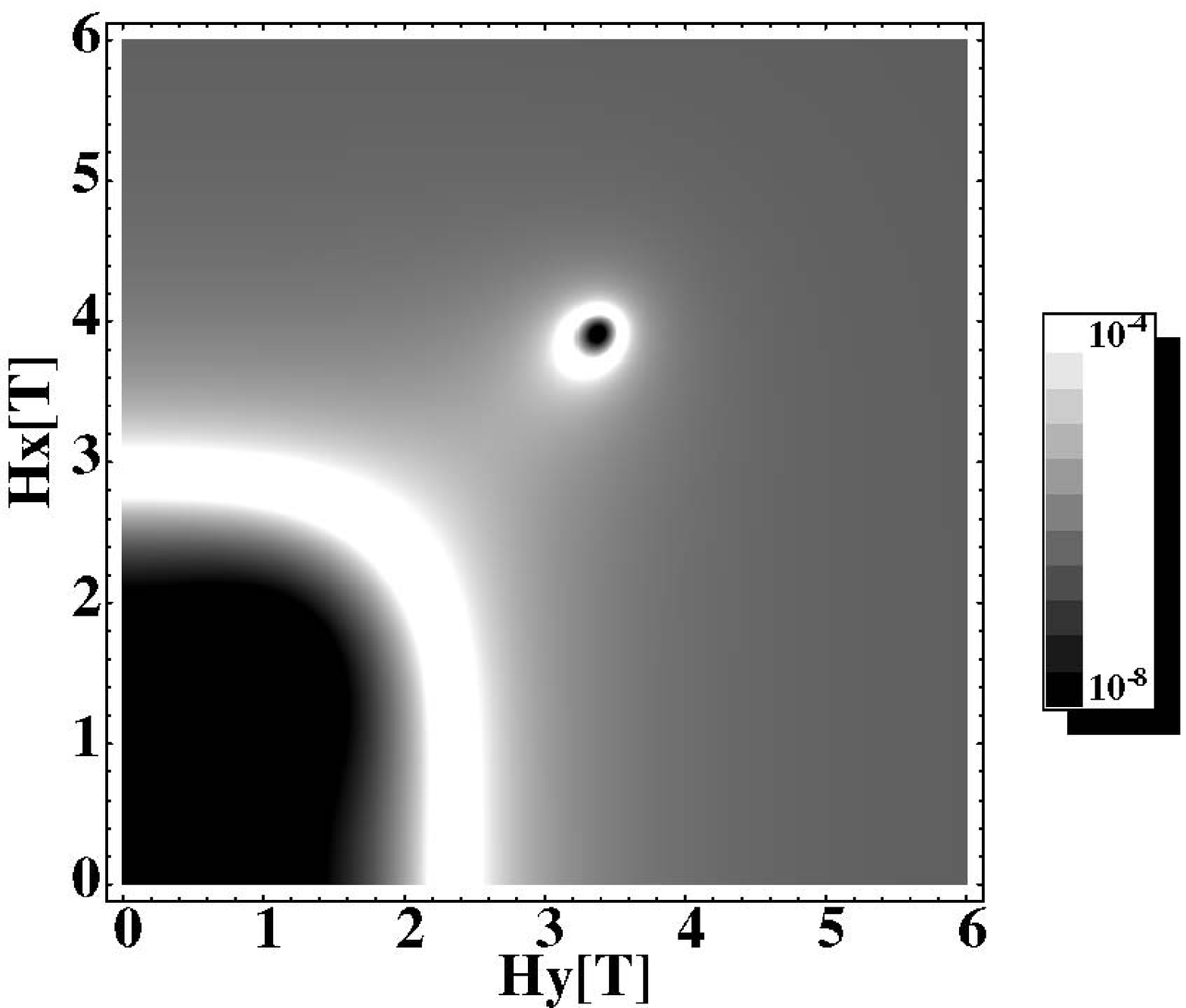}} &
      \resizebox{40mm}{!}{\includegraphics{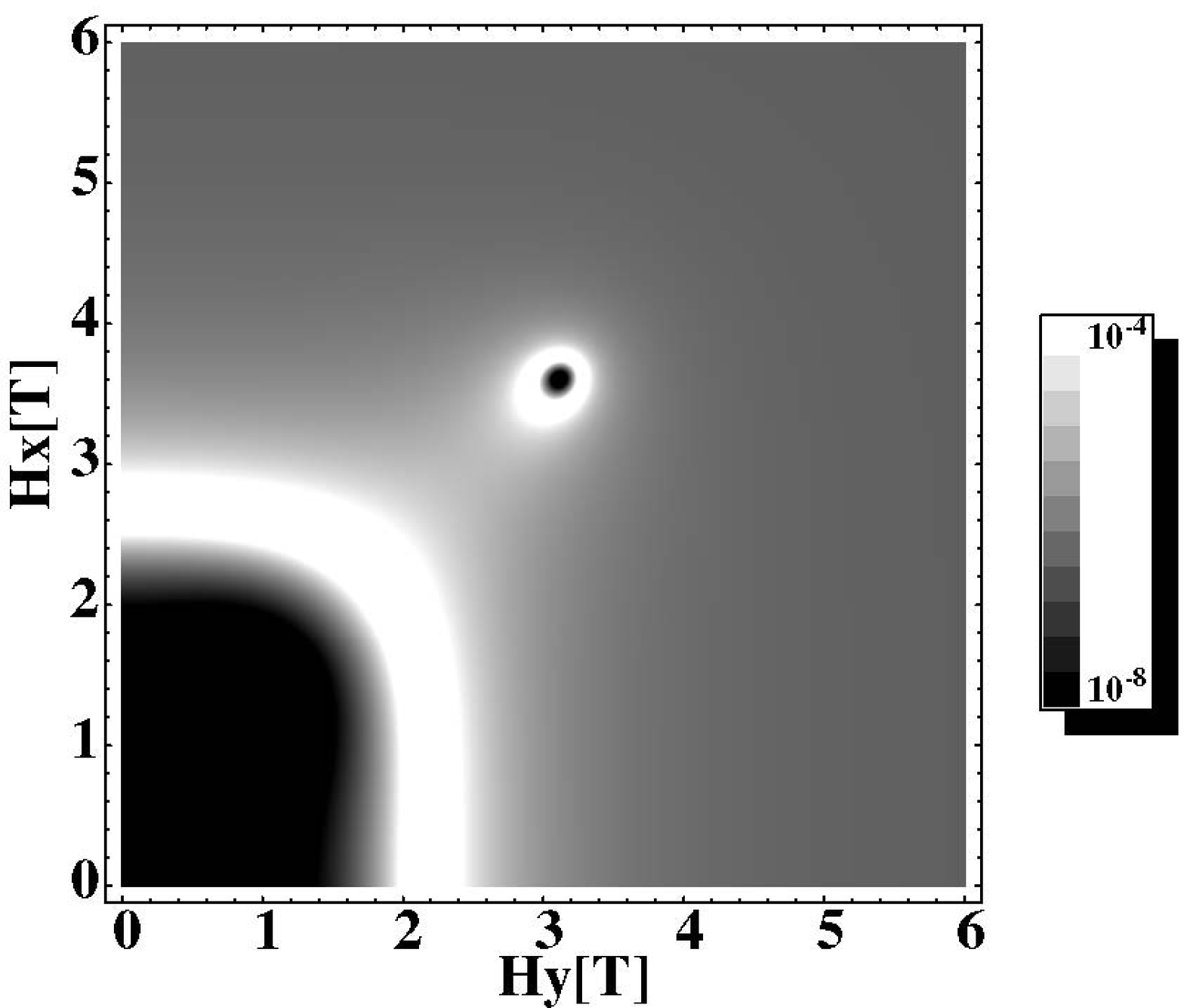}} \\
    \multicolumn{1}{c}{\mbox{\bf (c)}} &
		\multicolumn{1}{c}{\mbox{\bf (d)}} \\
      \resizebox{40mm}{!}{\includegraphics{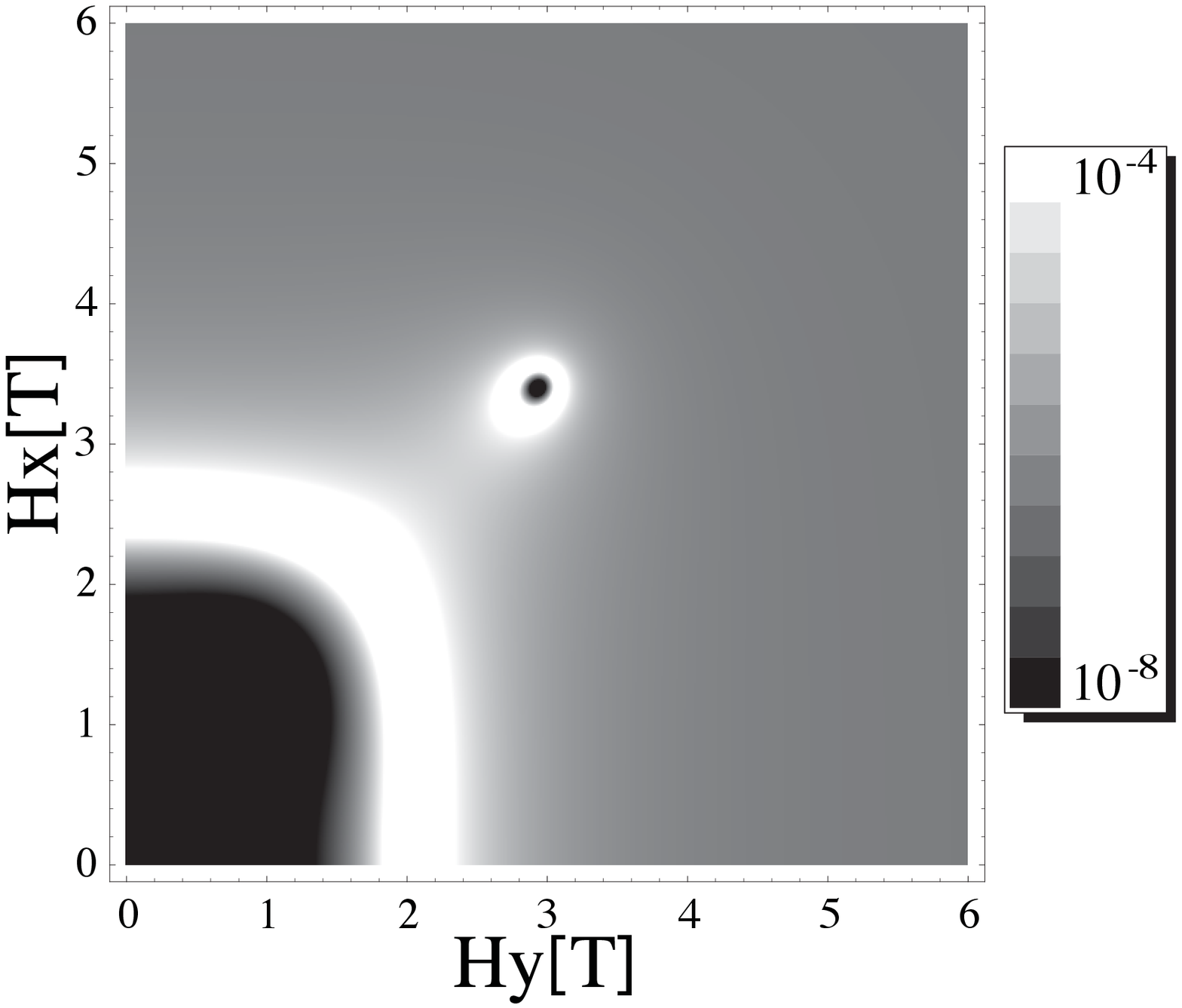}} &
      \resizebox{40mm}{!}{\includegraphics{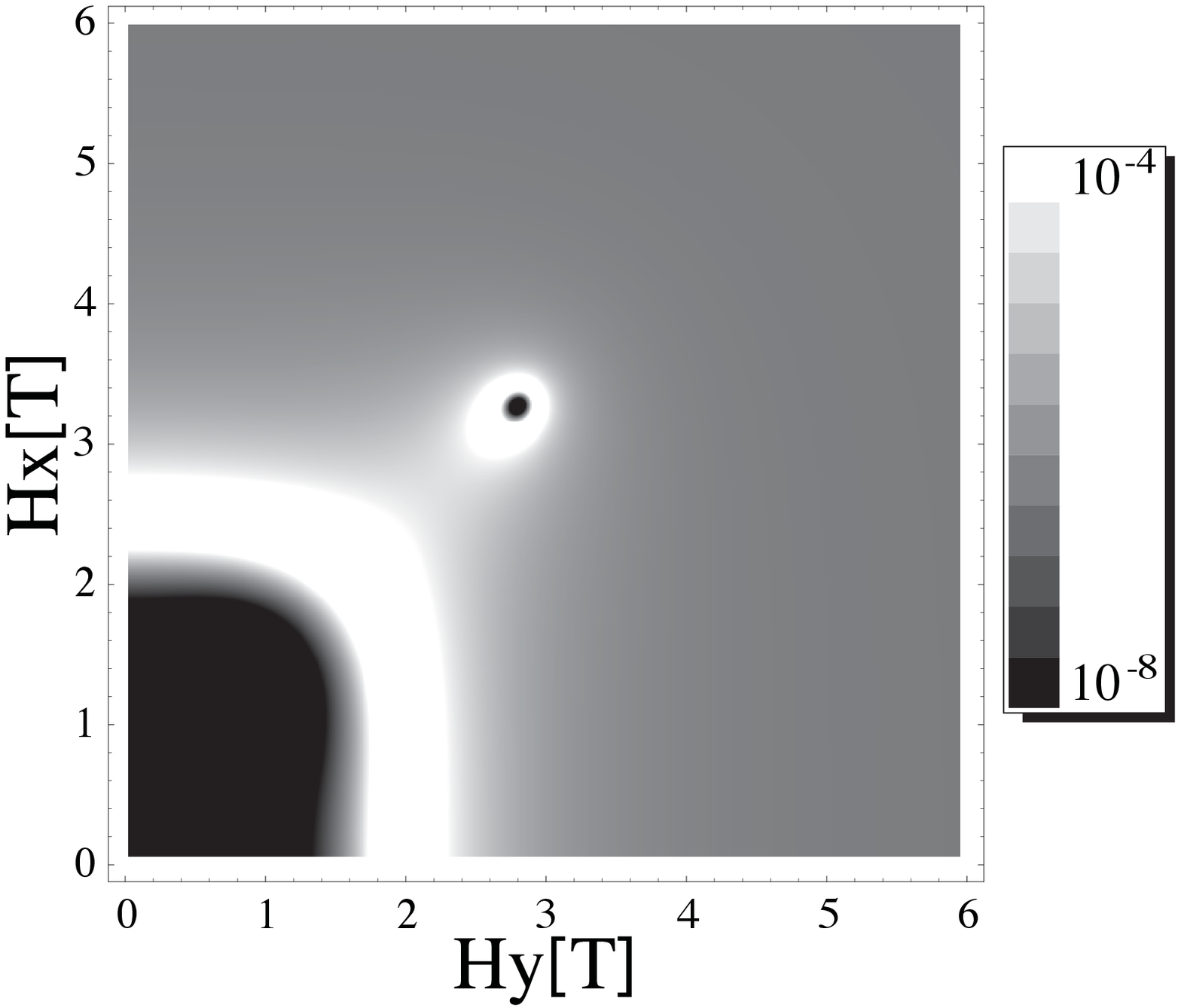}} \\
    \multicolumn{1}{c}{\mbox{\bf (e)}} &
		\multicolumn{1}{c}{\mbox{\bf (f)}} \\ 
    \end{tabular}
    \caption{Plots showing the temperature dependence of the
    conductance as a function of the transverse magnetic field for the
    tunnel splittings between states $|4\rangle$ and $|-4\rangle$ for
    the following temperature values: (a) $T/T_K=1.5$, (b)
    $T/T_K=1.6$, (c) $T/T_K=1.7$, (d) $T/T_K=1.8$, (e) $T/T_K=1.9$,
    and (f) $T/T_K=2.0$. We used a bias voltage of $V_b=V_L-V_R=2.7\mu
    V$.}
	\label{Conduct}
  \end{center}
\end{figure}

\begin{figure}[ht]
\includegraphics[width=8cm]{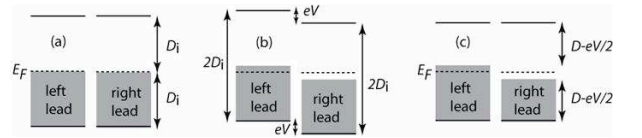}
\caption[]{Diagrams showing the asymmetric cut of the left- and
right-contact bands when a finite bias is applied.}
\label{Cutband}
\end{figure}

Consider the situation where one moves from zero point $b_\bot^{(n)}$
to the magnetic field value $b_\bot = b_\bot^{(l)} + \Delta b_\bot$,
where $\Delta b_\bot = \Delta h_\bot/\eta$. If $|eV| \ll
\Delta_0(b_\bot) \ll T_K$, the transmission amplitude is well
approximated by Eq. (\ref{amplitude}). On the other hand, for $|eV|
\gg T_K \gg \Delta_0(b_\bot)$, the transmission amplitude is given by
$A_{eV} = J^\perp_{eV}$. For the case $|eV|\sim\Delta_0(b_\bot) \gg
T_K$, we can expand $A_{\tilde{D} \approx \max\{T,\Delta_0(b_\bot)\}}$
up to second order in perturbation theory at the end of the flow,
yielding
\begin{eqnarray} 
A_{\tilde{D}}(\omega) & = & J^\perp_{\tilde{D}} + \rho_0
\int_{-\tilde{D}+eV/2}^{\tilde{D}-eV/2} d\epsilon^\prime
\frac{J^\perp_{\tilde{D}}\, J^z_{\tilde{D}}} {\omega-\epsilon^\prime}
\nonumber \\ & = & J^\perp_{\tilde{D}} + \rho_0 J^\perp_{\tilde{D}}
J^z_{\tilde{D}} \ln \left| \frac{\omega + \tilde{D} - eV/2} {\omega -
\tilde{D} + eV/2} \right|,
\label{conductance2}
\end{eqnarray}
where the integration limits account for the asymmetric cut of the
bands (see Fig. \ref{Cutband}). For $|eV|\sim\Delta_0(b_\bot) > T$,
the renormalization flow stops at
$\tilde{D}=\Delta_0(b_\bot)$. Substituting Eq. (\ref{conductance2})
into Eq. (\ref{conductance1}) and setting $E_F=0$, we obtain the
differential conductance up to third order in $J_{\tilde{D}}$ for both
positive and negative biases $eV$,
\begin{equation} 
\frac{G}{G_0} = \frac{\pi^2\rho_0^3}{16} J^{\perp 2}_{\tilde{D}}
 J^z_{\tilde{D}} \ln \left( \frac{\Delta_0} {\left| |eV| -
 \Delta_0 \right|} \right),
\label{conductance3}
\end{equation}
which agrees with the corresponding expression found in
Ref. \onlinecite{appelbaum}. Equation (\ref{conductance3}) represents
the conductance of the SMM as a function of temperature and tunnel
splittings between the $|4\rangle$ and $|-4\rangle$ states. In
Fig. \ref{Conduct} one can see how the conductance depends on the
temperature and takes a minimum value for those points where the
tunnel splitting is zero, which correspond to the zeros in
Fig. \ref{BPBT}. Note that our conductance formula depends only on the
bias voltage and not on the individual chemical potentials of the
leads, i.e. our conductance formula is gauge-invariant, which is a
result of the asymmetric band cutting shown in Fig.~\ref{Cutband}.

The two split Kondo peaks appear at $|eV|=\Delta_0(b_\bot)$.  Thus the
distance between the two peaks oscillates with the magnetic field,
following the renormalized periodic oscillations of the tunnel
splitting $\Delta_0(b_\bot)$.

\section{Conductance and the Kondo effect in a SMM in the cotunneling regime}
\label{sec:results1}

At zero temperature, the current in a single electron transistor can
be understood as a sequential process of single electrons tunneling in
and out of the left and right electrodes through the SMM, where the
transport channel is in between the electrochemical potentials of the
source and drain reservoir. The electron transport through the SMM is
also possible for any off-resonant energy, which is commonly called
the cotunneling regime, but it is expected to be very small compared
with the sequential tunneling. The cotunneling contribution can be
calculated by Fermi's golden rule in second-order perturbation theory,
i.e.
\begin{equation}
w_{i\rightarrow f}=\frac{2\pi}{\hbar}\left|\langle f | H | i \rangle +
\sum_{m}\frac{\langle f | H | m \rangle \langle m | H | i
\rangle}{E_m-E_f}\right|^2\rho(E),
\label{eqco1}
\end{equation}
and plays a dominant role whenever the sequential tunneling is
suppressed, i.e. $\left<f|H|i\right>=0$. In contrast to the Kondo
effect, where the tunnel coupling $t$ to the leads is very large, in
the cotunneling regime $t$ is typically small and thus it is
necessary to apply a gate voltage $V_g$ in order to get close to the
resonance condition of for example the $q=1$ charged state. Thus we
need to take only the electron scattering into account, thereby
neglecting the hole scattering contribution. Using the incoherent spin
states for temperatures around 1 K,\cite{gabriel} we can calculate the
cotunneling contribution by means of Eq. (\ref{eqco1}) in the
following form
\begin{eqnarray}
w_{\downarrow\rightarrow \uparrow} & = & \frac{2\pi}{\hbar}\left|
\frac{{_0}\langle \downarrow |{\cal H}_{\rm tot} | s \rangle \langle s
| {\cal H}_{\rm tot} |
\uparrow\rangle_{0}}{\tilde{U}/2-\Delta_1/2}\right. \nonumber\\ & &
+\left.\frac{{_0}\langle \downarrow | {\cal H}_{\rm tot} | a \rangle
\langle a | {\cal H}_{\rm tot} |
\uparrow\rangle_{0}}{\tilde{U}/2+\Delta_1/2} \right|^2 \rho(E),
\label{eqco2}
\end{eqnarray} 
where $\tilde{U}=U+V_g$.  Using Eq. (\ref{eqS}) in Eq. (\ref{eqco2})
we get
\begin{eqnarray}
w_{\downarrow\rightarrow \uparrow} & = & \frac{2\pi}{\hbar}\left|
\frac{{_0}\langle\downarrow |\tilde{\cal H} | \downarrow \rangle_{0}\,
{_0}\langle \uparrow | \tilde{\cal H} |
\uparrow\rangle_{0}}{\tilde{U}-\Delta_1}\right. \nonumber\\ & &
-\left.\frac{{_0}\langle \downarrow | \tilde{\cal H} | \downarrow
\rangle_{0}\, {_0}\langle \uparrow | \tilde{\cal H} |
\uparrow\rangle_{0}}{\tilde{U}+\Delta_1} \right|^2 \rho(E).
\label{eqco3}
\end{eqnarray}
After some simple algebra, we arrive at the following expression for
the tunnel rate process,
\begin{equation}
w_{\downarrow\rightarrow \uparrow} = \frac{2\pi}{\hbar} t^{4} \left|
\frac{2\Delta_1}{\tilde{U}^2-\Delta_{1}^2}\right|^{2} \rho(E),
\label{eqco4}
\end{equation}
where $t$ is the lead-molecule tunneling amplitude. Comparing
Eq. (\ref{eqco4}) with $J^\perp_{k,a;k^\prime,a^\prime}$ from
Eq. (\ref{eqj}) we see that the second contribution to Fermi's golden
rule is a function of the coupling constant, which is to be expected
since the Schrieffer-Wolff transformation is a perturbative approach
of the second order. The advantage of using the Schrieffer-Wolff
transformation is that you can apply the formalism of renormalization
theory to get a better description of the physical system near the
Kondo temperature.

\begin{figure}[ht]
\includegraphics[width=8cm]{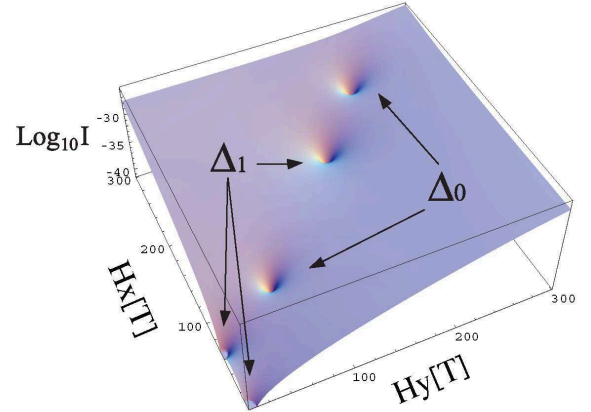}
\caption[]{Graph showing $\log_{10} I$ versus the transverse magnetic
field for $V_b=V_L-V_R=4\times 10^{-3}$eV, $V_g=0.01$eV and
$\gamma_{4,-4}=10^{12}$s$^{-1}$. One can see that the current is
suppressed at the zeros of the tunnel splittings $\Delta_1$ and
$\Delta_0$.}
\label{Cotunneling}
\end{figure}

Focusing on the Ni$_4$ single-molecule magnet, we use
Eq.~(\ref{eqco4}) to calculate the total cotunneling rate between
states $|\uparrow\rangle = |4\rangle$ and $|\downarrow\rangle =
|-4\rangle$ that will contribute to the current flowing through the
single-electron transistor,
\begin{equation}
W_{4,-4} = \frac{2\pi\, t^{4}\Delta_1^2}{\hbar} \int_{-eV/2-\tilde{U}}^{eV/2-\tilde{U}} \frac{dE}{(E^2-\Delta_1^2)^2},
\label{eqco5}
\end{equation}
where we integrate over all initial and final
states that are available within the range of the bias voltage
$V$. Performing the integration in Eq. (\ref{eqco5}) yields
\begin{eqnarray}
W_{4,-4} & = & \frac{4\pi\, t^{4}}{\hbar} \left[\frac{\tilde{U}-eV/2}{(\tilde{U}-eV/2)^2-\Delta_1^2}
-\frac{\tilde{U}+eV/2}{(\tilde{U}+eV/2)^2-\Delta_1^2} \right. \nonumber \\ & & \left. +
\frac{1}{2\Delta_1}\ln \left|\frac{(\Delta_1+eV/2)^2-\tilde{U}^2}{(\Delta_1-eV/2)^2-\tilde{U}^2}\right| \right].
\label{eqco6}
\end{eqnarray}
The total current flowing through the single-molecule magnet can be
calculated in terms of the density matrix by using the master
equation. Following the same procedure as in
Ref.~\onlinecite{gabriel}, we obtain the coupled differential
equations
\begin{eqnarray}
\dot{\rho}_{4} & = & \left( \frac{\Delta_1}{\hbar}\right)^2\frac{2\,
\gamma_{4,-4}} {V_g^2/\hbar^2+\gamma_{4, -4}^2}(\rho_{-4}-\rho_{4})
\nonumber \\ & & + W_{4,-4}\rho_{-4}-W_{-4, 4}\rho_{4},
\end{eqnarray}
and
\begin{eqnarray}
\dot{\rho}_{-4} & = & \left( \frac{\Delta_1}{\hbar}\right)^2\frac{2\,
\gamma_{4, -4}}{V_g^2/\hbar^2 + \gamma_{4,
-4}^2}(\rho_{4}-\rho_{-4})\nonumber \\ & & + W_{-4, 4}\rho_{4}-W_{4,
-4}\rho_{-4},
\end{eqnarray}
where $\gamma_{4, -4}$ is the incoherent tunneling rate from the lead
to the molecule. Solving the set of differential equations for
$\rho_{4}$ and $\rho_{-4}$ for the stationary case, we obtain the
current flowing through the SMM for the case where the source and the
drain leads are oppositely spin-polarized, as described in
Ref.~\onlinecite{gabriel}:

\begin{equation}
I = e\, W_{4, -4}\rho_{-4} = \frac{2\, e\, \gamma_{4,-4}\,
\Delta_1^2\, W_{4,-4}}{W_{4,-4}(V_g^2 + \gamma_{4,-4}^2\hbar^2) +
4\gamma_{4,-4}^2\Delta_1^2}.
\label{eqco7}
\end{equation}
Figure~\ref{Cotunneling} shows the cotunneling current as a function
of the transverse magnetic field. Interestingly, the current is
suppressed at the zeros of the tunnel splittings $\Delta_1$ and
$\Delta_0$, exactly as in the sequential tunneling
regime.\cite{gabriel}

\section{Conclusions}
\label{sec:conclusions}

The main contribution of this paper is to show how the total
Hamiltonian of a single-molecule magnet transistor can be mapped into
the Kondo Hamiltonian by means of a Schrieffer-Wolff
transformation. While the derivation of the effective Kondo
Hamiltonian in other contexts, such as quantum dots and ordinary
single (nonmagnetic) molecule coupled to leads is well-known, the case
is different for SMM. The dominant Kondo effect is unusual for a SMM,
since it involves a pseudospin of the molecule rather than its total
spin. We show that if the total spin of the molecule is reduced
(increased) for the charged states, the Kondo Hamiltonian exhibits an
antiferromagnetic (ferromagnetic) coupling, leading to screening
(anti-screening) of the total spin of the single-molecule magnet.  In
the case of antiferromagnetic coupling the renormalization leads to a
Kondo effect, i.e. the conductance through the SMM exhibits a
resonance at the Fermi energy.  In the case of the ferromagnetic
coupling the transverse exchange is renormalized to zero, in which
case the Kondo resonance is absent.  This result is in contrast to the
case of the Kondo effect in lateral quantum dots exhibiting only
antiferromagnetic exchange coupling, which is due to the fact that all
the spin states are degenerate in the absence of
anisotropies.\cite{Glazman2005} The standard Kondo screening of the
molecule magnetization by itinerant electrons in the leads is a very
weak effect in this context given the large spin of a SMM (its onset
is therefore likely to occur only at exceedingly small temperatures,
inaccessible to current experiments).

A careful derivation of the effective Kondo Hamiltonian shows that the
strong dependence of this phenomenon on the amplitude and orientation
of an external magnetic field. The strong uniaxial magnetic anisotropy
of the SMM combined with the weaker in-plane anisotropy creates a
pseudo spin 1/2 involving states with opposite magnetization
orientation. A transverse magnetic field modulates the tunnel barrier
between these states through a Berry phase interference effect. That,
in turn, modulates periodically the Kondo effect in SMMs.

We have calculated the conductance of the single-molecule transistor
in the presence of the Kondo effect by using the standard poor man's
scaling approach. We have show that in the case of antiferromagnetic
Kondo exchange coupling by applying a transverse magnetic field to a
single-molecule magnet with a large full- or half-integer spin $S>1/2$
it is possible to topologically induce or quench the Kondo effect of
the conductance of a current through the single-molecule magnet that
is sufficiently well coupled to metallic leads. We have also shown how
the zero points of the Berry phase oscillation become temperature
dependent above the Kondo temperature and how they change direction
within the plane (Fig. \ref{BPBT}). The latter indicates that the
parity of the Berry-phase oscillations
\cite{LossDelftGarg,Leuenberger_Berry} changes from an integer spin
$S=4$ to a half-integer spin $S=7/2$. We have also shown how this
motion affects the temperature dependence of the conductance (see
Fig. \ref{Conduct}). Interestingly, the maximum value of the
conductance encircles the zeros of the Berry phase oscillation,
providing a mechanism for establishing the location of these zeros
when the orientation of the molecule symmetry axis with respect to the
metallic contacts is not known. We illustrate these features of the
conductance of a SMM using as an example the new single-molecule
magnet Ni$_4$. In our view, due to its large ground state tunnel
splitting this is currently the best SMM available for the
experimental observation of the Berry-phase oscillations of the Kondo
resonance.

\section*{Acknowledgments}

The authors gratefully acknowledge useful discussions with George
Christou, Enrique del Barco, Leonid Glazman, Chris Ramsey, and Peter
Schmitteckert. E.R.M. acknowledges partial support through the NSF
Grant No. CCF 0523603 and by the I$^2$Lab at UCF. He also thanks the
Max-Planck Institute for the Physics of Complex Systems for its
hospitality. M.N.L. acknowledges partial support through the NSF Grant
No. ECCS 0725514.




\begin{thebibliography}{11}
\bibitem{obs1} The actual applied gate voltage is typically much
larger than $V_g$ due to screening and the setup geometry.
\bibitem{obs2} Generalizations to include additional charge sectors
are straightforward.
\bibitem{obs3} The state with $m_q=0$ can be easily included in this
description. However, since it typically requires a relatively high
temperature or bias to be accessed, it will be neglected here.
\end{thebibliography}
\end{document}